\documentclass[]{raa}            

\usepackage{graphicx,times}             
\usepackage[noblocks]{authblk}

\usepackage{cases}            

\usepackage{graphicx,times}

\usepackage{url}

\usepackage{pdflscape}

\usepackage{longtable}

\usepackage{multirow}
\begin{document}

   \title{The pc-scale radio structure of MIR-observed radio galaxies}

   \volnopage{Vol.0 (20xx) No.0, 000--000}      
   \setcounter{page}{1}          

   \author{Ye Yuan
      \inst{1,2,3}
   \and Minfeng Gu
      \inst{1,2}
   \and Yongjun Chen
      \inst{1,4}
   }

   \institute{Shanghai Astronomical Observatory, Chinese Academy of Sciences, Shanghai 200030, China; {\it yuany@shao.ac.cn}\\
        \and
             Key Laboratory for Research in Galaxies and Cosmology, Shanghai Astronomical Observatory, Chinese
             Academy of Sciences, 80 Nandan Road, Shanghai, 200030, China\\
        \and
            Uinversity of Chinese Academy of Sciences,19A Yuquan Road, Beijing 10049, China\\
        \and
            Key Laboratory of Radio Astronomy, Chinese Academy of Sciences, 2 West Beijing Road, Nanjing,
            JiangSu 210008, China\\
   }

   \date{Received~~2009 month day; accepted~~2009~~month day}

\abstract{ We investigated the relationship between the accretion process and jet properties by ultilizing the VLBA and mid-infrared (MIR) data for a sample of 45 3CRR radio galaxies selected with a flux density at 178 MHz $>16.4$ Jy, 5 GHz VLA core flux density $\geq$ 7 mJy, and MIR observations. The pc-scale radio structure at 5 GHz are presented by using our VLBA observations for 21 sources in February, 2016, the analysis on the archival data for 16 objects, and directly taking the measurements for 8 radio galaxies available in literatures. The accretion mode is constrained from the Eddington ratio with a dividing value of 0.01, which is estimated from the MIR-based bolometric luminosity and the black hole masses. While most FRII radio galaxies have higher Eddington ratio than FRIs, we found that there is indeed no single correspondence between the FR morphology and accretion mode with eight FRIIs at low accretion and two FRIs at high accretion rate. There is a significant correlation between the VLBA core luminosity at 5 GHz and the Eddington ratio. Various morphologies are found in our sample, including core only, single-sided core-jet, and two-sided core-jet structures. We found that the higher accretion rate may be more likely related with the core-jet structure, thus more extended jet. These results imply that the higher accretion rates are likely able to produce more powerful jets. There is a strong correlation between the MIR luminosity at 15 $\mu$m and VLBA 5 GHz core luminosity, in favour of the tight relation between the accretion disk and jets. In our sample, the core brightness temperature ranges from $10^{9}$ to $10^{13.38}$ K with a median value of $10^{11.09}$ K indicating that systematically the beaming effect may not be significant. The exceptional cases, FRIs at high and FRIIs at low accretion rate, are exclusively at the high and low end of the distribution of the flux ratio of VLBA core to 178 MHz flux density. It is not impossible that the location of these sources are due to the recent shining or weakening of the central engine (i.e., both accretion and jet).
}

   \authorrunning{Yuan et al. }            
   \titlerunning{Jets in radio galaxies }  

   \maketitle

%
%
\section{Introduction} \label{sec:intro}
It has been long found that the Fanaroff-Riley type I radio galaxies (FRIs) are edge-darkened, while Fanaroff-Riley type II radio galaxies (FRIIs) are edge-brightened (Fanaroff \& Riley \cite{1974MNRAS.167P..31F}). 
For a given host galaxy luminosity, FRIs have lower radio luminosities than FRIIs (Owen \& Ledlow \cite{1994ASPC...54..319O}). 
The primary reason for this difference is still not clear. 
There are two scenarios to explain this difference, due to either the different physical conditions in ambient medium (Gopal-Krishna \& Wiita \cite{2000A&A...363..507G}), or the difference in central engines, i.e., different accretion modes and/or jet formation processes (Ghisellini \& Celotti \cite{2001A&A...379L...1G}). 

About three decades ago, two different types of central engine were realized based on the analysis on powerful radio sources. 
Many powerful objects have strong optical and ultraviolet continua, for which one invokes copious and radiatively efficient accretion flows (quasars and broad line radio galaxies) (Begelman et al. \cite{1984RvMP...56..255B}). 
But many double radio sources, e.g. Cygnus A, lack this radiative signature, which can be instead perhaps explained by Blandford $\&$ Znajek (\cite{1977MNRAS.179..433B}) with a mechanism for electromagnetic extraction of rotational energy of black hole.

Later on, using spectropolarimetric observations, the quasar spectra were discovered in many radio galaxies (Antonucci \cite{1984ApJ...278..499A}), suggesting that all radio galaxies and radio quasars are powered by similar central engine. 
While the hidden quasars were detected in many radio galaxies, in some cases, they were not (Singal \cite{1993MNRAS.262L..27S}). Quasars hidden by dusty gas will re-radiate their absorbed energy in the infrared, therefore, the extensive observations with the infrared spectroscopy were made to search more robustly for the hidden quasars in radio galaxies (e.g., Ogle et al. \cite{2006ApJ...647..161O}; Haas et al. \cite{2005A&A...442L..39H}; Leipski et al. \cite{2009ApJ...701..891L}, etc). The targets were selected by their diffuse radio flux density, to minimize any orientation biases. The Spitzer observations indicate that there are two types of central engine for radio galaxies, which do not show a correlation exactly with FR class. 
Ogle et al (2006) showed that about half of the narrow-line FRII radio galaxies have a mid-IR luminosity
at 15 $\rm \mu m$ of $\rm > 8\times10^{43}erg~s^{-1}$, indicating strong thermal emission from hot dust in the active galactic
nucleus, just like the matched quasars. However, they also found that another half do not. These MIR-weak sources do not contain a powerful accretion disk, and they may be fit with nonthermal, jet-dominated AGNs, where the jet is powered by a radiatively inefficient accretion flow or black hole spin-energy, rather than energy extracted from an accretion disk. 
The dismatch with FR class was also found in FRIs. Leipski et al. (2009) reported most FRIs lack powerful type-1 AGN, but it is not tenable to generalize on associations between FRI galaxies and nonthermal only AGNs, and a fraction of FRIs do have warm dust emission which could be attributed to hidden type-1 nuclei (Antonucci \cite{2002apsp.conf..151A}).

The different central engine types in radio galaxies could be constrained from their IR luminosty. 
On the scales of the relativisitic jets being produced, the central engine may deeply differ, which call the investigations to understand how accretion mode affects the innermost radio emission. Very long baseline interferometry (VLBI) is one of the most poweful tools to detect the jet properties at pc-scales. 
In this work, we combine the VLBA and mid-infrared observations for a sample of radio galaxies, to study the relation of accretion process and jet properties at pc-scale.
Our sample is shown in Section \ref{sec:2}, and Section \ref{sec:3} is for the VLBA and MIR data. We present the results and discussions in Section \ref{sec:4}, while the conclusion is provided in Section \ref{sec:5}.  
Throughout the paper, we use a cosmology with $\rm H_0 = 70~km~ s^{-1}\rm Mpc^{-1}$,  $\Omega_{\rm m} = 0.30$, $\Omega_\Lambda = 0.70$. The spectral index $\alpha$ is defined as $f_{\nu}\propto\nu^{\alpha}$, in which $f_{\nu}$ is the flux desity at frequency $\nu$.

\section{Sample} \label{sec:2}
To systematically study the relationship between the accretion mode and the pc-scale jet properties, we choose a sample from 3CRR\footnote{http://3crr.extragalactic.info/} catalogue (Laing et al. \cite{1983MNRAS.204..151L}). 
There are 173 sources in the 3CRR catalogue, including 43 quasars, 10 broad-line radio galaxies, and 120 narrow-line radio galaxies. The original 3CRR catalogue has a flux limit of 10 Jy at 178 MHz, and is the canonical low-frequency selected catalogue of bright radio sources. From 3CRR sample, the MIR observations have been well studied for a well-defined, radio flux-limited sample of 50 radio galaxies with a flux density at $178$ MHz $>16.4$ Jy, and 5 GHz VLA core flux density $\geq$ 7 mJy (e.g. Ogle et al. \cite{2006ApJ...647..161O}; Haas et al. \cite{2005A&A...442L..39H}; Leipski et al. \cite{2009ApJ...701..891L}, etc). The MIR emission enable us to explore the existence of hidden quasars, thus we use this subsample in our study. We carefully searched the VLBI observation for all these 50 objects, and found that 27 sources have already been observed with VLBA. We observed the ramaining 23 targets with VLBA at 5 GHz. In two objects, the poor $uv$ data preclude us to make good images. Moreover, in three of the 27 sources with VLBA observation from archive, the VLBA data are not useful to make final images.  After excluding these five sources, 
the final sample consisits of 45 radio galaxies with MIR detections and VLBA observations either by us or from archive.
The essential information of the sample are list in Table \ref{tab_1}, in which 30 sources belong to FRIIs, 11 sources FRIs, and the remaining 4 sources are core-dominated source (Laing et al. \cite{1983MNRAS.204..151L}). 

\begin{center}

	\begin{footnotesize}
		\setlength{\tabcolsep}{2pt}
		\begin{longtable}{lcccccccccccr}
			\caption{Sample of 3CRR radio galaxies.\label{tab_1}}\\
			\hline\hline
			name &alias &ID & 
			$z$ & FR &log $M_{\rm BH}$ & D &
			$f_{\textrm{VLA}}$ & $f_{\textrm{178}}$& 
			$f_{\textrm{MIR}}$ &log $L_{\textrm{bol}}$ &
			Calibrator & Distance 
			\\ 
			& & &  &  & ($\rm M_{\odot}$)& ($\rm Mpc$)&
			($\rm mJy$) & ($\rm Jy$) & ($\rm mJy$)  & ($\rm erg $~$s^{-1}$) &  & ($\rm deg$)  \\
			(1) & (2) & (3)& (4) & 
			(5) & (6) & (7) &(8)&(9)& (10)& (11) &(12)&(13)\\
			\endfirsthead
			\caption{Continue.}\\
			\hline\hline
			name &alias &ID & 
			$z$ & FR &log $M_{\rm BH}$ & D &
			$f_{\textrm{VLA}}$ & $f_{\textrm{178}}$& 
			$f_{\textrm{MIR}}$ &log $L_{\textrm{bol}}$ &
			Calibrator & Distance 
			\\ 
			& & &  &  & ($\rm M_{\odot}$)& ($\rm Mpc$)&
			($\rm mJy$) & ($\rm Jy$) & ($\rm mJy$)  & ($\rm erg $~$s^{-1}$) &  & ($\rm deg$)  \\
			(1) & (2) & (3)& (4) & 
			(5) & (6) & (7) &(8)&(9)& (10)& (11) &(12)&(13)\\
			\hline
			\endhead
			\hline
			\caption{continued on next page}
			\endfoot
			\hline
			\endlastfoot			 
			\hline
			3C 31  & UGC00689 & Bo015 & 0.0167 & I  & 8.65 & 72.4  & 92    & 18.3  & 17.19 &43.82&  &  \\ 
			3C 33 & & BG239 & 0.0595 & II & 8.50 & 266   & 24    & 59.3  & 75 & 45.20&  & \\
			3C 47 & & BG239 & 0.425 & II & 9.20 & 2330   & 73.6  & 28.8  & 34.39 &46.32 & &\\
			3C 48$^c$ & & $...$     & 0.367 & C & 8.80 & 1960 & 896   & 60    & 110.91 &46.62&  &  \\
			3C 66B & UGC01841 & Bo015 & 0.0215 & I  & 8.58 & 93.6 & 182   & 26.8  & 4.76 & 43.56& &   \\
			3C 79 & & BG239 & 0.2559 & II & 8.80 &  1294   & 10    & 33.2  & 42.08 & 46.03& 0316+162 & 2.23 \\
			3C 84$^b$  && $...$      & 0.0177 & I &  8.89 & 76.8   & 59600 & 66.8  & 1146.04&45.30 &  &  \\
			3C 98 & 0356+10 & BG158 & 0.0306 & II & 8.21 & 134  & 9     & 51.4  & 48.8$^a$ &44.28 & &  \\
			3C 109 && BT065 & 0.3056 & II & 9.30 & 1586    & 263   & 23.5  & 120.02 & 46.51& &  \\
			3C 123$^b$ && $...$      & 0.2177 & II & 7.87 & 1078   & 100   & 206   & 2.8 &44.99 & &   \\
			3C 138& & BC081 & 0.759 & C & 8.70 &  4700  & 94    & 24.2  & 15.1$^a$ & 46.12& &  \\
			3C 147$^b$ && $...$     & 0.545 & C & 8.70 & 3142   & 2500  & 65.9  & 22.4$^a$ &46.03&  &  \\
			3C 173.1 && BG239 & 0.292 & II & 8.96 &  1510  & 7.4   & 16.8  & 0.6 & 44.67&0708+742 & 0.78  \\
			3C 192 & &BG239 & 0.0598 & II & 8.43 & 268   & 8     & 23    & 3.2 & 44.13& 0759+252 &1.19 \\
			3C 196 & &BG239 & 0.871 & II & 9.60 & 5570  & 7     & 74.3  & 22.9 &  46.68&0804+499 &1.82  \\
			3C 208 & &BH167 & 1.109 & II & 9.40 & 7510  & 51    & 18.3  & 5.8 & 46.38& &  \\
			3C 212 & &BH057 & 1.049 & II & 9.20 & 7010 & 150   & 16.5  & 15.5 &46.68 & &  \\
			3C 216$^b$& &$...$      & 0.668 & II & 7.00 & 4020 & 1050  & 22    & 28.7 &46.58&  &  \\
			3C 219& & BG239 & 0.1744 & II & 8.77 & 841.7  & 51    & 44.9  & 11.2 & 45.31& &  \\
			3C 220.1 && BG239 & 0.61  & II & 8.40 & 3600  & 25    & 17.2  & 2.4 & 45.66&&    \\
			3C 226 && BG239 & 0.82  & II & 8.05 & 5200 & 7.5   & 16.4  & 15.65 & 46.52&0943+105 &0.77 \\
			3C 228 && BG239 & 0.5524 & II & 8.27 & 3194   & 13.3  & 23.8  & 0.99 & 45.29& 0951+175& 3.15 \\
			3C 234 && BG239 & 0.1848 & II & 8.88 & 897.5  & 90    & 34.2  & 239  &46.39&  & \\
			3C 254 & &BG239 & 0.734 & II & 9.30 & 4510 & 19    & 21.7  & 11.6$^a$ &46.01&  & \\
			3C 263 && BG239 & 0.652 & II & 9.10 & 3910  & 157   & 16.6  & 29.8 &46.57& &   \\
			3C 264 && BK125 & 0.0208 & I & 8.57 & 90.5  & 200   & 28.3  & 10.32 & 43.80& &  \\
			3C 272.1$^b$ && $...$     & 0.0029 & I & 8.40 & 12   & 180   & 21.1  & 27.6&42.76 &  &  \\
			3C 274 & J1230+12 & W040  & 0.0041 & I & 8.86 & 18   & 4000  & 1144.5 & 42.96 &43.19&  &  \\
			3C 275.1& & BG239 & 0.557 & II & 8.30 & 3230  & 130   & 19.9  & 8.4 & 46.03& &  \\
			3C 286$^b$ & & $...$       & 0.849 & C & 8.50 & 5400   & 5554  & 27.3  & 7.64$^a$ &45.97&  &  \\
			3C 288 && BG239 & 0.246 & I  & 9.50 & 1240  & 30    & 20.6  & 0.6 & 44.55& &  \\
			3C 300 && BG239 & 0.272 & II & 8.49 & 1390 & 9     & 19.5  & 0.7 &  44.67&1417+172&2.61 \\
			3C 309.1 & J1459+71 &BB233 & 0.904 & II & 9.10 & 5830 & 2350  & 24.7  & 17.2$^a$&46.29 &  &  \\
			3C 326 & 1549 & BG202 & 0.0895 & II & 8.23 & 409 & 13    & 22.2  & 0.39 &43.69&  &   \\
			3C 338 && BV017 & 0.0303 & I & 9.07 & 133  & 105   & 51.1  & 2.4 &  43.56&&   \\
			3C 380 && BM157 & 0.691 & I  & 9.40 & 4190 & 7447  & 64.7  & 40.4 & 46.72&&  \\
			3C 382 && BT065 & 0.0578 & II & 8.75 & 258  & 188   & 21.7  & 114 & 45.33& &  \\
			3C 386 && BG239 & 0.0177 & I & 8.57 & 76.8  & 120   & 26.1  & 2.47 & 43.20& &  \\
			3C 388 && BG239 & 0.0908 & II & 8.81 & 415  & 62    & 26.8  & 0.84 & 43.96& &  \\
			3C 390.3$^b$& & $...$      & 0.0569 & II & 8.92 &  254  & 330   & 51.8  & 164 & 45.44& &  \\
			3C 401 && BG239 & 0.201 & II & 9.18 & 986  & 32    & 22.8  & 0.8 & 44.50& &  \\
			3C 436 && BG239 & 0.2145 & II & 8.66 & 1060  & 19    & 19.4  & 1.5 & 44.76& &   \\
			3C 438 & &BG239 & 0.29  & II & 8.74 &  1490 & 7.1   & 48.7  & 0.45 & 44.56& 2202+363&2.23 \\
			3C 452 & &BB199 & 0.0811 & II & 8.46 & 369  & 130   & 59.3  & 45 &45.24 & &   \\
			3C 465 && V018  & 0.0293 & I  & 8.77 & 128  & 270   & 41.2  & 3.17 & 43.63& &  \\	
		\end{longtable}	
	\end{footnotesize}
\end{center}
{\footnotesize
	Notes: Columns (1) - (2): source name and alias name; Column (3): VLBA project code, $^b,^c$ - VLBA measurements from Fomalont et al. (\cite{2000ApJS..131...95F}), and Worrall et al. (\cite{2004MNRAS.347..632W}), respectively; Columns (4) and (5): redshift and FR types I and II, C represents the core-dominated source; Column (6): black hole mass; Column (7): luminosity distance;  Columns (8) - (11): the VLA core flux density at 5 GHz, and the 178 MHz flux density, and mid-infrared flux desity at 15 $\mu$m ($^a$ - at 24 $\mu$m), and the bolometric luminosity; Columns (12) - (13): phase calibrators for phase-reference observations, and its separation to the source.}

\section{Data compilation} \label{sec:3}
In this work, the VLBA and MIR data are essential to study the relationship between the accretion mode and pc-scale jets in radio galaxies, which are complied from our observations and archive data.

\subsection{VLBA observations and data reduction} \label{subsec:Vdata}

The VLBA observations of our sample consists of three groups. In the first group, we performed VLBA observations at C-band with a total observing time of 20 hours for 23 sources in three blocks for scheduling convenience on Feb. 13, 14, and 15, 2016 (program ID: BG239). In two of these 23 sources, we are not able to make images due to poor $uv$ data quality, thus this group finally consists of 21 objects. 
Among these 21 sources, thirteen radio galaxies can be self-calibrated with observing time of 30 mins for each target, while for the remaining eight sources, the phase-referencing is required with on-source time of 40 mins individually. 
These sources and the related phase calibrators are list in Table \ref{tab_1}. 
Group two has 16 radio galaxies, of which the VLBA observational data can be downloaded from NRAO archive\footnote{https://archive.nrao.edu/archive/advquery.jsp} (see program ID in Table \ref{tab_1}). For the rest eight sources, the third group,  the measurements of jet components can be directly obtained from literatures (Fomalont et al. \cite{2000ApJS..131...95F}; Worrall et al.  \cite{2004MNRAS.347..632W}). 

The data reduction was performed for the sources in groups one and two. Data are processed with AIPS in a standard way. Before fringe fitting, we correct for Earth orientation,  remove dispersive delay from ionosphere,  and calibrate the amplitude
by using system temperature and gain curve. Phase calibration is followed in order by correcting for instrumental phase and delay 
using pulse-calibration data, removing the residual phase, delay and rate for relative strong targets  by fringe fitting on source
itself. For weaker targets, phase-referencing technique is taken by applying the residual phase, delay and rate solutions 
from phase-referencing calibrator to the corresponding  target in interpolating method.
Imaging and model-fitting were performed in DIFMAP and
the final results are given in Table \ref{tab_2}, in which the measurements of jet components directly adopted from literatures are also given for eight sources. Tentatively, we assume the brightest component to be radio core in this work. 
The VLBA radio images for each object are shown in Figure \ref{fig:bg239} and Figure \ref{fig:other}, for groups one and two, respectively. All images are at 5 GHz, except for 3C 208, in which 8 GHz data is used since there is no 5 GHz data available.

\begin{center}
	\begin{footnotesize}
		\begin{longtable}{lcccccccr}
			\caption{Results for the radio galaxies \label{tab_2}}\\
			\hline\hline
			Name & Comps.  & FR & 
			$S$ & $r$ & $\theta$ &
			$a$ & $b/a$ &log $T_{\rm B}$ \\ 
			& &  & (mJy) & (mas) & (deg)&
			(mas) &  &(K) \\
			(1) & (2) & (3) & (4) & 
			(5) & (6) & (7)&(8) & (9) \\
			\endfirsthead
			\caption{Continue.}\\
			\hline\hline
			Name & Comps.  & FR & 
			$S$ & $r$ & $\theta$ &
			$a$ & $b/a$ &log $T_{\rm B}$ \\ 
			& &  & (mJy) & (mas) & (deg)&
			(mas) &  &(K) \\
			(1) & (2) & (3) & (4) & 
			(5) & (6) & (7)&(8) & (9) \\
			\hline
			\endhead
			\hline
			\caption{continued on next page}
			\endfoot
			\hline
			\endlastfoot			 
			\hline
			3C 31   & C  & I  &80.37  & 0.24  & 178.16  & 0.22  & 1.00  &  11.09  \\
			&    &   & 15.97  & 0.85  & $-19.75$  & 0.14  & 1.00  &         \\
			&    &   & 3.09  & 7.79  & $-14.43$  & 1.55  & 1.00  &         \\
			&    &   & 5.02  & 3.00  & $-13.66$  & 1.18  & 1.00  &         \\
			&    &   & 1.16  & 11.45  & $-14.64$  & 1.40  & 1.00  &         \\
			3C 33   & C  & II  & 20.89  & 0.07  & $-149.68$  & 0.24  & 1.00   & 10.46  \\
			&    &   & 1.90   & 4.57  & $-155.97$  & 0.25  & 1.00   &        \\
			&    &   & 12.94  & 0.26  & 25.18      & 14.18 & 0.05   &        \\
			&    &   & 2.65   & 15.41 & $-158.56$  & 5.34  & 0.32   &        \\
			&    &   & 1.24   & 12.15 & 25.19      & 0.65  & 1.00   &        \\
			&    &   & 1.69   & 42.71 & 26.19      & 5.78  & 0.33   &        \\
			&    &   & 0.91   & 16.14 & 21.54      & 1.11  & 1.00   &        \\
			&    &   & 0.59   & 35.32 & 25.76      & 0.07  & 1.00   &        \\
			&    &   & 0.60   & 54.06 & 26.83      & 1.77  & 1.00   &        \\
			&    &   & 0.41   & 4.14  & 27.21      & 0.49  & 1.00   &        \\
			3C 47   & C  & II  & 50.97  & 0.05  & 33.95  & 0.14  & 1.00  & 11.57 \\
			&    &   & 5.52  & 2.06  & $-149.00$  & 0.24  & 1.00  &         \\
			&    &   & 5.13  & 7.46  & $-149.60$  & 13.71  & 0.08  &         \\
			&    &   & 0.78  & 20.88  & $-149.04$  & 1.09  & 1.00  &        \\
			&    &   & 1.31  & 4.49  & $-146.83$  & 0.41  & 1.00  &         \\
			&    &   & 1.04  & 12.30  & $-145.93$  & 0.75  & 1.00  &         \\
			&    &   & 1.10  & 1.58  & 22.11  & 0.42  & 1.00  &         \\
			&    &   & 0.80  & 7.01  & $-153.43$  & 0.48  & 1.00  &         \\
			&    &   & 0.24  & 25.96  & $-147.85$  & 0.62  & 1.00  &         \\
			3C 48   & C  & C  &56.10  &       & 171.00  & 2.20  & 0.18  &  9.93  \\
			3C 66B  & C  &I   &137.32  & 0.31  & $-124.58$  & 0.03  & 1.00  &  13.05  \\
			&    &   & 84.78  & 0.55  & 56.66  & 0.03  & 1.00  &        \\
			&    &   & 1.52  & 21.79  & 53.79  & 1.24  & 1.00  &         \\
			&    &   & 16.05  & 2.43  & 59.46  & 0.36  & 1.00  &        \\
			&    &   & 5.38  & 4.80  & 55.70  & 0.16  & 1.00  &         \\
			&    &   & 21.33  & 11.02  & 53.64  & 13.38  & 0.10  &        \\
			&    &   & 0.66  & 7.23  & 57.92  & 0.41  & 1.00  &        \\
			3C 79   & C  & II  &27.10  & 0.02  & 108.07  & 0.46  & 1.00  &  10.16  \\
			&    &   & 0.98  & 2.41  & $-72.30$  & 1.97  & 1.00  &         \\
			&    &   & 0.86  & 7.06  & $-71.87$  & 2.50  & 1.00  &         \\
			&    &   & 0.32  & 14.89  & $-13.89$  & 0.12  & 1.00  &         \\
			&    &   & 0.30  & 24.28  & $-48.36$  & 0.16  & 1.00  &         \\
			3C 84   & C  & I  &17752.00 &       & 154.00  & 4.60  & 0.78  &  10.90  \\
			&    &   & 5833.00  &       & 170.00  & 6.60  & 0.14  &         \\
			&    &   & 3084.00  &       & 161.00  & 5.70  & 0.28  &        \\
			3C 98   & C  & II  &44.87  & 0.01  & 14.63  & 0.21  & 1.00  &  10.88 \\
			3C 109  & C  & II  & 221.48  & 0.05  & $-7.57$  & 0.22  & 1.00  &  11.75  \\
			&    &   & 28.37  & 1.91  & 155.42  & 0.11  & 1.00  &         \\
			&    &   & 6.30  & 8.66  & 150.16  & 1.51  & 1.00  &         \\
			&    &   & 9.16  & 3.66  & 154.24  & 1.05  & 1.00  &         \\
			&    &   & 3.72  & 26.09 & 147.10  & 1.25  & 1.00  &         \\
			&    &   & 5.40  & 13.83  & 152.72  & 2.92  & 1.00  &         \\
			&    &   & 3.77  & 20.73  & 149.93  & 2.51  & 1.00  &         \\
			&    &   & 4.47  & 5.37  & 151.00  & 1.01  & 1.00  &         \\
			&    &   & 4.40  & 10.81  & 150.35  & 1.43  & 1.00  &         \\
			3C 123  & C  & II  &111.00  &       & 92.00  & 3.90  & 0.77  &  9.00  \\
			3C 138  & C  &  C & 130.47  & 0.04  & $-113.69$  & 0.60  & 1.00  &  10.90       \\ 
			&    &   &76.55  & 1.64  & $-109.71$  & 0.16  & 1.00  &   \\
			&    &   & 47.50  & 3.94  & 112.07  & 3.51  &0.21  &         \\
			&    &   & 98.36  & 6.30  & $-92.35$  & 0.19  & 1.00  &         \\
			&    &   & 22.17  & 10.83 & 58.89  & 6.77  & 0.25 &         \\
			&    &   & 38.20  & 14.58  & 27.92  & 8.06  & 1.00  &         \\
			&    &   & 10.93  & 19.00 & 6.22  & 5.11  & 0.26 &         \\
			&    &   & 20.80  & 2.69  & $-199.01$  & 0.54  & 1.00  &         \\
			3C 147  & C  &  C &882.00  &       & 171.00  & 2.10  & 0.57  &  10.77  \\
			&    &   & 506.00  &       & 16.00  & 2.80  & 0.18  &         \\
			&    &   & 676.00  &       & 146.00  & 5.00  & 0.28  &         \\
			&    &   & 222.00  &       &       & 1.40  & 0.43  &         \\
			3C 173.1& C  & II  &14.81  &0.02  & $-178.81$  & 0.06  & 1.00  &  11.69  \\
			3C 192  & C  &  II &13.63  & 0.02  & $-150.73$  & 0.03  & 1.00  &  12.09  \\
			3C 196  & C  & II  &14.25  & 0.02  & $-69.89$  & 0.18  & 1.00  &  11.03 \\
			3C 208$^a$ & C &II & 72.62  & 0.02  & 170.04  & 0.15  & 0.88  & 11.65  \\
			&    &   & 9.01  & 3.80  & $-96.16$  & 0.90  & 0.45  &         \\
			&    &   & 4.54  & 0.78  & $-95.66$  & 0.07  & 1.00  &         \\
			3C 212  & C  & II  &118.40  & 0.04  & 125.47  & 0.16  & 1.00  &  12.14  \\
			&    &   & 12.87  & 1.06  & $-40.29$  & 0.07  & 1.00  &         \\
			&    &   & 3.41  & 11.64  & $-35.84$  & 1.63  & 1.00  &        \\
			&    &   & 10.79  & 2.53  & $-43.52$  & 0.84  & 1.00  &         \\
			&    &   & 1.40  & 15.20 & $-36.59$  & 0.17  & 1.00  &         \\
			&    &   & 1.00  & 17.67  & $-35.88$  & 0.24  & 1.00  &         \\
			3C 216  & C  & II  &620.00  &       & 152.00  & 0.80  & 0.38  &  11.70  \\
			&    &   & 85.00  &       & 149.00  & 2.70  & 0.22  &         \\
			3C 219  & C  & II  &44.47  & 0.10  & 42.98  & 0.09  & 1.00  &  11.73  \\
			&    &   & 12.08  & 1.30  & $-137.30$  & 2.83  & 0.04  &         \\
			&    &   & 0.78  & 7.24  & $-141.25$  & 4.03  & 0.21  &         \\
			&    &   & 0.24  & 15.82  & $-140.63$  & 1.66  & 1.00  &         \\
			3C 220.1 & C &II   &22.26  & 0.16  & $-66.26$  & 0.58  & 0.32  &  10.35  \\
			&    &   & 6.57  & 0.68  & 80.44  & 1.31  & 0.20  &         \\
			&    &   & 0.64  & 5.43  & 82.56  & 3.08  & 0.82  &         \\
			&    &   & 0.31  & 11.14  & 107.56  & 0.06  & 1.00  &         \\
			3C 226  & C  & II  &17.22  & 0.03  & $-11.04$  & 0.03  & 1.00  &  12.66  \\
			&    &   & 0.25  & 1.34  & 148.61  & 0.17  & 1.00  &         \\
			&    &   & 0.27  & 1.50  & $-25.26$  & 0.13  & 1.00  &         \\
			3C 228  & C  & II  &18.41  & 0.03  & 1.36  & 0.05  & 1.00  &  12.10  \\
			&    &   & 0.90  & 2.20  & $-173.66$  & 0.09  & 1.00  &         \\
			&    &   & 0.30  & 9.45  & $-167.45$  & 0.36  & 1.00  &         \\
			&    &   & 0.25  & 4.34  & $-164.92$  & 0.26  & 1.00  &         \\
			3C 234  & C  & II  &19.71  & 0.19  & $-166.70$  & 0.28  & 1.00  &  10.39  \\
			&    &   & 12.48  & 5.19 & 65.34  & 1.52  & 0.39  &         \\
			&    &   & 1.87  & 8.49  & 67.38  & 0.62  & 1.00  &         \\
			&    &   & 10.44  & 1.53  & 67.37  & 0.28  & 1.00  &         \\
			&    &   & 0.47  & 17.01  & 65.87  & 1.56  & 1.00  &         \\
			&    &   & 0.23  & 5.79  & $-108.60$  & 0.22  & 1.00  &         \\
			&    &   & 0.20  & 29.54  & 66.67  & 0.84  & 1.00  &         \\
			&    &   & 0.23  & 11.87  & 63.40  & 0.17  & 1.00  &         \\
			&    &   & 0.23  & 13.55  & 72.18  & 0.20  & 1.00  &         \\
			&    &   & 0.16  & 3.93  & 39.60  & 1.03  & 1.00  &         \\
			3C 254  & C  & II  &18.19  & 0.11  & 86.87  & 0.02  & 1.00  &  12.99  \\
			&    &   & 3.27  & 1.25  & $-71.87$  & 0.04  & 1.00  &         \\
			&    &   & 2.61  & 3.20  & $-71.01$  & 1.06  & 1.00  &         \\
			&    &   & 0.71  & 5.98  & $-66.40$  & 0.80  & 1.00  &         \\
			&    &   & 0.32  & 9.71 & $-70.11$  & 1.09  & 1.00  &         \\
			&    &   & 0.26  & 12.27 & $-72.26$  & 0.24  & 1.00  &         \\
			3C 263  & C  & II  &111.60  & 0.10 & $-72.97$  & 0.01  & 1.00  &  13.38  \\
			&    &   & 49.77  & 0.91  & 108.43  & 2.09  & 0.18  &         \\
			&    &   & 4.06  & 3.15  & 111.08  & 0.06  & 1.00  &         \\
			&    &   & 10.64  &       & 113.84  & 3.31  & 0.16  &         \\
			&    &   & 0.36  &       & 109.23  & 0.72  & 1.00  &         \\
			&    &   & 1.69  &       & 111.88  & 14.05  & 0.08  &         \\
			3C 264  & C  &I   &159.07  & 0.08  & 174.75  & 0.13  & 1.00  &  11.84  \\
			&    &   & 20.00  & 1.55  & 28.35  & 0.35  & 1.00  &        \\
			&    &   & 18.15  & 4.39  & 24.31  & 1.91  & 1.00  &         \\
			3C 272.1 & C & I  &187.00  &       &       & 1.10  & 0.64  &  10.23  \\
			&    &   & 13.00  &       &       & 2.00  & 1.00  &         \\
			3C 274  & C  &I   &850.85  & 0.14  & 84.64  & 0.40  & 1.00  &  11.58  \\
			&    &   & 390.81  & 0.89  & $-82.20$  & 0.63  & 1.00  &         \\
			&    &   & 308.21  & 0.89  & 98.28  & 0.46  & 1.00  &         \\
			&    &   & 43.08  & 2.00  & 107.23  & 1.17  & 1.00  &         \\
			&    &   & 137.06  & 2.05  & -83.38  & 0.29  & 1.00  &         \\
			3C 275.1 & C & II  &262.91  & 0.19  & 144.58  & 0.72  & 0.08  &  12.03  \\
			&    &   & 77.80  & 1.38  & $-25.10$  & 0.44  & 1.00  &         \\
			&    &   & 6.56  & 8.06  & $-18.32$  & 1.68  & 1.00  &        \\
			&    &   & 7.16  & 3.16 & $-15.05$  & 1.22  & 1.00  &         \\
			&    &   & 2.02  & 14.96  & $-19.14$  & 2.70  & 1.00  &        \\
			3C 286  & C  &  C &1723.00  &       & 33.00  & 4.60  & 0.78  &  10.41  \\
			&    &   & 978.00  &       & 61.00  & 7.40  & 0.50  &         \\
			&    &   & 192.00  &       & 108.00  & 2.60  & 0.58  &         \\
			3C 288  & C  & I  &20.18  & 0.20  & 74.08  & 0.26  & 1.00  &  10.52  \\
			3C 300  & C  & II  &25.12  & 0.02  & 106.06  & 0.50  & 0.10  &  11.06  \\
			3C 309.1 & C & II  &287.32  & 0.38  & $-36.12$  & 0.16  & 1.00  &  12.46  \\
			&    &   & 325.52  & 23.84  & 163.32  & 1.42  & 1.00  &         \\
			&    &   & 83.38  & 24.70  & 165.76  & 0.82  & 1.00  &         \\
			&    &   & 283.09  & 1.01  & 162.64  & 0.35  & 1.00  &         \\
			&    &   & 148.12  & 22.44  & 167.73  & 2.24  & 1.00  &         \\
			&    &   & 242.09  & 50.30 & 155.79  & 14.43  & 1.00  &         \\
			&    &   & 142.48  & 40.73  & 163.31  & 4.42  & 1.00  &         \\
			&    &   & 6.38  & 52.10 & 145.74  & 0.69  & 1.00  &         \\
			&    &   & 35.80  & 2.54  & 165.33  & 0.36  & 1.00  &         \\
			&    &   & 16.38  & 36.63  & 174.28  & 1.80  & 1.00  &         \\
			&    &   & 11.38  & 31.04  & 164.18  & 0.42  & 1.00  &        \\
			3C 326  & C  & II  &35.75  & 0.04  & 133.83  & 0.04  & 1.00  &  12.28  \\
			&    &   & 0.86  & 4.55  & $-37.95$  & 0.59  & 1.00  &         \\
			&    &   & 1.22  & 0.84  & $-66.60$  & 0.28  & 1.00  &         \\
			&    &   & 0.27  & 2.69  & 130.60  & 0.55  & 1.00  &         \\
			3C 338  & C  & I  &90.81  & 0.12  & $-113.55$  & 0.16  & 1.00  &  11.42  \\
			&    &   & 33.52  & 1.50  & 79.75  & 1.24  & 1.00  &         \\
			&    &   & 13.40  & 3.00  & $-94.21$  & 0.54  & 1.00  &         \\		
			&    &   & 10.04  & 6.02  & 78.11  & 1.45  & 1.00  &         \\
			&    &   & 7.34  & 10.46  & $-90.40$  & 2.15  & 1.00  &         \\
			&    &   & 5.50 & 10.33  & 86.11  & 1.26  & 1.00  &         \\
			&    &   & 7.25  & 20.47 & 91.12  & 2.66  & 1.00  &         \\
			3C 380  & C  & I  &1038.99  & 0.61  & 149.70  & 1.37  & 0.15  &  11.87  \\
			&    &   & 240.58  & 9.09  & $-31.75$  & 1.78  & 0.42  &         \\
			&    &   & 153.21  & 2.87  & $-26.57$  & 1.91  & 0.31  &         \\
			3C 382  & C  & II  &120.30  & 0.43  & $-125.88$  & 0.06  & 1.00  &  12.42  \\
			&    &   & 104.67  & 0.56  & 52.66  & 0.31  & 1.00  &         \\
			&    &   & 22.94  & 1.99  & 56.80  & 0.75  & 1.00  &         \\
			3C 386  & C  & I  &17.19  & 0.05  & $-136.79$  & 0.33  & 1.00  &  10.07  \\
			3C 388  & C  & II  &35.55  & 0.16  & 48.61  & 0.19  & 1.00  &  10.92  \\
			&    &   & 6.11  & 1.38  & $-114.17$  & 0.16  & 1.00  &        \\
			&    &   & 2.70  & 3.81   & $-118.78$ & 1.38  & 1.00  &         \\
			&    &   & 0.77  & 9.74  & $-117.36$  & 1.59  & 1.00  &         \\
			&    &   & 1.37  & 2.45  & 54.78  & 2.42  & 1.00  &         \\
			&    &   & 0.33  & 15.07  & $-124.03$  & 0.40  & 1.00  &         \\
			&    &   & 0.16  & 6.68  & $-118.28$  & 0.40  & 1.00  &         \\
			3C 390.3 & C & II  &463.00  &       & 159.00  & 1.90  & 0.21  &  10.68  \\
			&    &   & 261.00  &       & 166.00      & 1.40  & 0.28  &        \\
			3C 401  & C  & II  &20.94  & 0.10  & 36.42  & 0.21  & 1.00  &  10.69  \\
			&    &   & 5.58  & 0.62  & $-169.43$  & 0.43  & 1.00  &         \\
			&    &   & 0.49  & 5.08  & $-157.05$  & 0.18  & 1.00  &         \\
			&    &   & 0.24  & 17.04  & $-164.52$  & 1.46  & 1.00  &         \\
			&    &   & 2.06  & 1.93  & $-156.70$  & 0.44  & 1.00  &         \\
			3C 436  & C  & II  &16.79  & 0.62  & 3.21  & 0.70  & 0.43  &  9.92  \\
			&    &   & 0.88  & 5.36  & $-3.76$  & 0.90  & 1.00  &         \\
			&    &   & 0.39  & 15.27  & $-28.93$  & 0.21  & 1.00  &         \\
			&    &   & 0.27  & 36.68  & $-16.86$  & 0.78  & 1.00  &         \\
			3C 438  & C  & II  &16.00  & 0.03  & $-106.51$  & 0.11  & 1.00  &  11.19  \\
			3C 452  & C  & II  &95.31  & 0.26  & $-95.22$  & 0.84  & 1.00  & 10.04  \\
			&    &   & 74.44  & 2.12 & 85.89  & 1.20  & 1.00  &         \\
			&    &   & 27.54  & 1.90  & $-89.84$  & 0.54  & 1.00  &         \\
			&    &   & 9.76  & 8.73 & $-88.61$  & 1.32  & 1.00  &         \\
			&    &   & 8.06  & 4.99  & $-87.66$  & 2.52  & 1.00  &        \\
			&    &   & 5.64  & 9.43  & 81.59  & 1.26  & 1.00  &        \\
			3C 465  & C  &I   &90.03  & 0.49  & 135.88  & 0.30  & 1.00  &  10.87  \\
			&    &   & 29.16  & 1.54  & $-59.42$  & 0.81  & 1.00  &         \\
			&    &   & 12.22  & 3.12  & $-56.69$  & 0.56  & 1.00  &         \\
			&    &   & 3.66  & 8.11 & $-49.55$  & 0.67  & 1.00  &         \\
			&    &   & 7.58  & 4.45  & $-56.07$  & 0.86  & 1.00  &         \\						
		\end{longtable}	
	\end{footnotesize}
\end{center}
{\footnotesize
	Notes: Column (1): source name, $^a$ - at 8 GHz; Column (2): components, C represents the radio core; Column (3): FR types I and II, C represents the core-dominated source; Column (4): flux density; Columns (5) - (6): component position, and its position angle; Column (7): major axis; Column (8): axial ratio; Column (9): brightness temperature.}

\subsection{Mid-infrared data} \label{subsec:Mdata}
We collected the MIR data for our sample from NASA/IPAC Extragalactic Database (NED)\footnote{http://ned.ipac.caltech.edu/}, which is originally from the observations using either Spitzer IRS/MIPS or ISOCAM (e.g. Ogle et al. \cite{2006ApJ...647..161O}; Haas et al. \cite{2005A&A...442L..39H}; Leipski et al. \cite{2009ApJ...701..891L}; Temi et al. \cite{2005ApJ...622..235T}). In the sample, the flux density at 15 $\mu$m are available for 39 sources, and only 24 $\mu$m flux density are obtained in the remaining six radio galaxies (i.e., 3C 98, 3C 254, 3C 309.1, 3C 138, 3C 147, and 3C 286, see Table \ref{tab_1}).

\section{RESULTS AND DISCUSSION} \label{sec:4}
\subsection{Accretion mode} \label{sec:4.1}
FRIs have lower radio luminosities than FRIIs for a similar host galaxy luminosity (Owen \& Ledlow \cite{1994ASPC...54..319O}).
FRIs and FRIIs have shown clear dividing line in the radio and optical luminosity plane, which can be re-expressed as a line of constant ratio of the jet or the disk accretion power with the Eddington luminosity. 
This implies the accretion process plays a more important role in FRIs and FRIIs dichotomy than a different environment (Ghisellini \& Celotti \cite{2001A&A...379L...1G}).

Quasars hidden by dusty gas will re-radiate their absorbed energy in the infrared.  Ogle et al. (\cite{2006ApJ...647..161O}) investigated the MIR emission using the Spitzer survey of 3C objects, including radio galaxies and quasars, selected by the relatively isotropic lobe emission. 
They argued that most of the MIR-weak sources may not contain a powerful accretion disk. 
It is likely that in the nonthermal, jet-dominated AGNs, the jet is powered by a radiatively inefficient accretion flow or black hole spin-energy, rather than energy from accrtion disk. 
Two different central engines are recognized for FRIs or FRIIs in their study, with the dividing value of the luminosity at 15 $\mu$m of $\rm 8\times10^{43} ~ergs ~s^{-1}$. The sources with the luminosity above it are suggested to contain a radiatively efficient accretion flow.

Instead of a fixed dividing luminosity,  the accretion mode is investigated from the Eddington ratio $L_{\rm bol}/L_{\rm Edd}$ in this work, in which $L_{\rm bol}$ and $L_{\rm Edd}$ are the bolometric and Eddington luminosities, respectively.
The black hole masses of 17 sources are collected from various literatures (McLure et al. \cite{2006NewAR..50..782M}; Wu \cite{2009MNRAS.398.1905W};  McNamara et al. \cite{2011ApJ...727...39M}; Mingo et al. \cite{2014MNRAS.440..269M}). For the rest 28 radio galaxies, the black hole masses were estimated by using the relationship between the host galaxy absolute magnitude at R band ($M_{\rm R}$) and black hole mass provided by McLure et al. (\cite{2004MNRAS.351..347M}), 
\begin{equation}
\rm log~\it (\frac{\rm M_{\rm BH}}{\rm M_{\odot}})=\rm -0.5\it M_{\rm R}\rm -2.74
\end{equation}
in which, the $M_{\rm R}$ was calculated from the $R$ magnitude in the updated online 3CRR catalogue.  

In this work, the bolometric luminosity $L_{\rm bol}$ is calculated from mid-infrared luminosity either at 15 or at 24 $\mu$m, using the relation in 
Runnoe et al. (\cite{2012MNRAS.426.2677R}),
\begin{equation}
\rm log~\it L_{\rm bol}=\rm (10.514\pm4.390)+(0.787\pm0.098)~log({\nu}\it L_{\rm \nu,15\mu m})
\end{equation}
\begin{equation}
\rm log~\it L_{\rm bol}=\rm (15.035\pm4.766)+(0.688\pm0.106)~log({\nu}\it L_{\rm \nu,24\mu m})
\end{equation}
in which, a spectral indice of $\alpha_{\nu}=-1$ is used for k-correction.

We adopted a conventional value of $L_{\rm bol}/L_{\rm Edd}= 10^{-2}$  to separate radiatively efficient or inefficient accretion mode (e.g., Hickox et al. \cite{2009ApJ...696..891H}).  
The relationship between the VLBA core luminosity at 5 GHz and the Eddington ratio is presented in Figure \ref{fig:F_vlba}. The rest frame 5 GHz luminosity is estimated from the VLBA 5 GHz or 8 GHz (for 3C 208) core flux density using a spectral indice of $\alpha=0$. While most FRII radio galaxies have higher Eddington ratio than FRIs, we found that there is indeed no single correspondence between the FR morphology and accretion mode. The eight out of thirty FRIIs ($26.7\%$) may have low accretion rate with $L_{\rm bol}/L_{\rm Edd}< 10^{-2}$, and the rest 22 objects ($73.3\%$) are at high accretion mode with $L_{\rm bol}/L_{\rm Edd}\ge 10^{-2}$. In contrast, two out of eleven FRIs ($18.2\%$), and $81.8\%$ FRIs are at radiatively efficient and inefficient accretion mode, respectively. There is a significant correlation between the VLBA core luminosity at 5 GHz and the Eddington ratio, with a Spearman correlation coefficient of $r=0.820$ at $\gg 99.99$ per cent confidence. This implies that the higher accretion rate are likely able to produce more powerful jets.

The correlation between the MIR luminosity at 15 $\mu$m and VLBA 5 GHz core luminosity is also investigated in Figure \ref{fig:MIR_core}. The luminosity at 15 $\mu$m in six sources were estimated from 24 $\mu$m using a spectral indice of $\alpha_{\nu}=-1$. A significant correlation is found between two parameters with a Spearman correlation coefficient of $r=0.849$  at $\gg 99.99$ per cent confidence. After excluding the common dependence on redshift, the partial Spearman rank correlation method (Macklin \cite{1982MNRAS.199.1119M}) shows that the significant correlation is still present with a correlation coefficient of $r=0.635$ at $\gg 99.99$ per cent confidence. The linear fit gives 
\begin{equation}
\rm log~\it L_{\rm core,5GHz}=\rm (0.951\pm0.083)~log({\nu}\it L_{\rm \nu,15\mu m})\rm -(0.263\pm3.655)
\end{equation}
While in the flux-limited low-frequency radio survey like 3CRR sample, the low-frequency emission is mostly dominated by the lobe, which, however, is normally located at the jet end, thus represents the past jet activity. In contrast, the MIR, and especially the pc-scale VLBA core emission are instantaneously and comtemporarily from the central engine.   
The strong correlation strongly indicates the tight relation between the accretion disk and jets, as found in various works (e.g., Cao \& Jiang\cite{1999MNRAS.307..802C}; Gu et al. \cite{2009MNRAS.396..984G}).

In the framework of unification scheme of AGNs, FRIs are unified with BL Lac objects (BL Lacs), and FRIIs with flat-spectrum radio quasars (FSRQs) (Antonucci \cite{1993ARA&A..31..473A}; Urry \& Padovani \cite{1995PASP..107..803U}). The blazars consists of BL Lacs and FSRQs, and are characteristic of strong beaming effect due to the jets pointing towards us with small viewing angles. The jets in FSQRs are found to have stronger power and higher velocity than those in BL Lacs (e.g., Gu et al. \cite{2009MNRAS.396..984G}; Chen \cite{2018arXiv180305715C}). On the other hand, the Eddington ratios of BL Lacs are systematically lower than those of radio quasars with a rough division at $L_{\rm bol}/L_{\rm Edd} \sim 0.01$, which imply that the accretion mode of BL Lacs may be different from that of radio quasars (e.g., Xu et al. \cite{2009ApJ...694L.107X}). The radio galaxies used in this study has its own advantages in avoiding the strong contaminaion of jet beaming effect on the VLBA core emission, since the jet viewing angle are usually large in radio galaxies. Our results of higher accretion rate likely associated with stronger jet are generally in agreement with the unification scheme.

\subsection{Pc-scale Radio Morphology} \label{sec:4.2}

It can be clearly seen from the high-resolution VLBA 5 GHz images in Figures \ref{fig:bg239} and \ref{fig:other} that there are various morphologies in our sample sources, including 10 core only, 29 one-sided core-jet, and 6 two-sided core-jet structures. The two-sided core-jet structure is found in 3C 33, 3C 38, 3C 338, 3C 452 (see Figures \ref{fig:bg239} and \ref{fig:other}), 3C 147, and 3C 286 (Fomalont et al. \cite{2000ApJS..131...95F}). In this work, we will not distinguish the latter two categories, instead we call them all as core-jet structure.
The radio morphologies was further studied with the source fraction of the specified structure in 17 radido galaxies with inefficient accretion flow and 28 efficient ones.
At low Eddington ratio ($<10^{-2}$), we found that six out of seventeen ($35.3\%$) exhibit core only structure, and the remaining sources ($64.7\%$) have core-jet morphology. In contrast, core only and core-jet present in 3 ($10.7\%$) and 25 ($89.3\%$) sources, respectively, at high Eddington ratio ($\ge10^{-2}$). It thus seems that the higher accretion rate may be more likely related with the core-jet structure. For a similar distribution of viewing angles likely presents in our sample of radio galaxies, radio morphology perhaps can reflect some jet information like strength and speed in different accretion models.  
A core-jet radio morphology likely indicates the source jet moving at higher speed with relatively powerful. 
However, a naked core may indicates a relatively weaker jet with lower speed. 
Based on our analysis, we found that the radiatively inefficient accretion flow may perhaps be also inefficient in producing powerful jets moving at lower speed, while the radiatively efficient one shows higher probability on forming strong jets with higher speed. This is consistent with the correlation between the VLBA core luminosity at 5 GHz and the Eddington ratio shown in Figure \ref{fig:F_vlba}. In a broader framework, this is also consistent with the radio-quiet populations. LINERs and Seyferts can be analog to two accretion systems (Kewley et al. \cite{2006MNRAS.372..961K}). LINERs seem to have radio cores more optically thick than those of Seyferts, and their radio emission is mainly confined to a compact core or base of a jet, thus it is likely that the radiatviely inefficient accretion flow is likely to host a more compact VLBI pc-scale core, than that of radiative efficient one. 

The pc-scale VLBA projected linear size $l$ of sources is estimated as the largest distance among radio components for core-jet sources, while directly as the major axis for core-only galaxies.
The distribution of pc-scale VLBA size for all sources is presented in Figure \ref{fig:size}, except for eight objects, in which the size is not available in literatures. There is a broad range with most sources in 1 - 100 pc, and the jet extends to about 300 - 400 pc in several core-jet objects. We find a significant correlation between the linear size and the Eddington ratio with a correlation coefficient of $r=0.671$ at $\gg 99.99$ per cent confidence (see Figure \ref{fig:size}). This indicates that the higher accretion rate may have more extended jet, again supporting our results of more powerfully jets in higher-accretion system.

\subsection{Brightness Temperature} \label{sec:4.3}

From the high-resolution VLBA images, the brightness temperature of radio core $T_{\rm B}$ in the rest frame can be estimated with (Ghisellini et al. \cite{1993ApJ...407...65G})
\begin{equation}
T_{\rm B}=1.77\times10^{12}(1+z)(\frac{S_{\nu}}{\rm Jy})(\frac{\nu}{\rm
	GHz})^{-2}(\frac{\theta_{\rm d}}{\rm mas})^{-2} ~~\rm K
\end{equation} 
in which $z$ is the redshift, $S_{\nu}$ is core flux desity at frequency $\nu$, and $\theta_{\rm d}$ is the angular diameter, $\theta_{\rm d}$ = $\sqrt{ab}$ with $a$ and $b$ being the major and minor axes, respectively.  
There is an important parameter Doppler factor $\delta$, which can be restricted by 
\begin{equation}
\delta=T_{\rm B}/T_{\rm B}^{'}
\end{equation} 
in which $T_{\rm B}^{'}$ is the intrinsic brightness temperature. 
The core brightness temperature distribution diagram is presented in Figure \ref{fig:T_B}. 
In our sample, the core brightness temperature ranges from $10^{9}$ to $10^{13.38}$ K with a median value of $10^{11.09}$ K (see also in Table \ref{tab_2}). Most sources are in the range of $10^{10} - 10^{12}$ K, less than the inverse Compton catastrophic limits $10^{12}$ K (Kovalev et al. \cite{2005AJ....130.2473K}). Therefore, systematically the beaming effect may not be significant in our sample, although it may not be trivial in some cases, for example, in 3C 263, the source with the highest $T_{\rm B}$. In comparison, the VLBA core brightness temperatures of blazars typically range between $10^{11}$ and $10^{13}$ K with a median value near $10^{12}$ K, and can even extend up to $5\times10^{13}$ K (Kovalev et al. \cite{2005AJ....130.2473K}, \cite{2009ApJ...696L..17K}). These results are basically in agreement in the framework of unification scheme of AGNs, with FRIs/FRIIs and BL Lacs/FSRQs. The strong beaming effect results in high brightness temperature of the radio cores in blazars, while it is less pronounced in radio galaxies because of large jet viewing angles.

We have analyzed the correlation of the brightness temperature and the Eddington ratio in Figure \ref{fig:T_B}. There is no correlation between two parameters, and the distribution of $T_{\rm B}$ is similar at high and low accreiton rate.

\subsection{Compared with VLA data} \label{sec:4.4}

We collected VLA 5 GHz flux density for our sources from 3CRR catalogue, then we compared the VLBA with VLA flux density. The flux ratio of VLBA core to VLA core, and ratio of VLBA total to VLA core, are plotted with the Eddington ratio in Figure \ref{fig:compac}. The flux ratio between VLBA and VLA can in principle give information on the source compactness, since they represent the source structure at different scales, with normally the former at pc-scale, and the latter at kpc-scale. There are no correlations between the flux ratio and the Eddington ratio. The flux ratio covers more than one order of magnitude, and there is no systematical difference between the high and low accretion regimes. It's interesting to see that the VLBA core flux density is higher than VLA core in many sources. This can be most likely due to variability. This is even more pronounced when considering the VLBA total flux density. In this case, the VLBA total flux is higher than VLA core in majority of objects, implying the variability may be common in our sample.

\subsection{Core/lobe Flux Density Ratio} \label{sec:4.5}

In comparison of VLBI pc-scale core flux desity with 178 MHz flux density, we would investigate the present status of core radio activity. 
It might be possible that those sources with weak MIR dust emission are just recently at radiatively inefficient accretion model, while the large scale radio morphology was produced by past radiatively efficient accretion model. 
Therefore, their core/lobe flux density ratio are expected to be low. 
In previous works (e.g., Ogle et al. \cite{2006ApJ...647..161O}), the core and lobe luminosity ratio is indeed less in MIR-weak FRIIs than in MIR-luminous FRIIs at VLA. 

The ratio of VLBA core to 178 MHz flux density is plotted with the MIR luminosity and the Eddington ratio in Figure \ref{fig:corelobe}. While a MIR luminosity at 15 $\rm \mu m$ of $\rm 8\times 10^{43}~ erg~ s^{-1}$ is adopted to distinguish the MIR-weak and MIR-luminous sources in Ogle el al. (\cite{2006ApJ...647..161O}), we further use the Eddington ratio in recognizing the accretion mode. The flux ratio of VLBA core to 178 MHz covers about two orders of magnitude, and there is no single dependence of the flux ratio on the either the Eddington ratio or MIR luminosity. Similar behaviours are seen in the panels of the flux ratio with MIR luminosity and Eddington ratio. Considering solely the high and low accretion rate regime, there is no correlation between the radio flux ratio and MIR luminosity/Eddington ratio. The distribution of the flux ratio at high accretion rate is broader than that at low rate, which mainly concentrated on lower flux ratio and does not extend to very high values.   Interestingly, the FRIIs with low MIR luminosity (below $\rm 8\times 10^{43}~ erg~ s^{-1}$) or low accretion rate ($L_{\rm bol}/L_{\rm Edd}< 10^{-2}$) are exclusively at the lower end of the distribution of radio flux ratio. In contrast, two MIR-luminous or highly accreting FRIs are all at high end. It is not impossible that the location of these sources are due to the recent shining or weakening of the central engine (i.e., both accretion and jet), resulting a higher or lower VLBA core luminosity, thus a lower or higher flux ratio of VLBA core to 178 MHz.

\section{SUMMARY} \label{sec:5}
We investigated the role of the accretion model in creating the VLBI jets by ultilizing the VLBA and MIR data for a sample of 45 3CRR radio galaxies. The accretion mode is constrained from the Eddington ratio, which is estimated from the MIR-based bolometric luminosity and the black hole masses. While most FRII radio galaxies have higher Eddington ratio than FRIs, we found that there is indeed no single correspondence between the FR morphology and accretion mode with eight FRIIs at low accretion and two FRIs at high accretion rate. There is a significant correlation between the VLBA core luminosity at 5 GHz and the Eddington ratio. We found that the higher accretion rate may be more likely related with the core-jet structure, thus more extended jet. These results imply that the higher accretion rate are likely able to produce more powerful jets. There is a strong correlation between the MIR luminosity at 15 $\mu$m and VLBA 5 GHz core luminosity, in favour of the tight relation between the accretion disk and jets. In our sample, the core brightness temperature ranges from $10^{9}$ to $10^{13.38}$ K with a median value of $10^{11.09}$ K indicating that systematically the beaming effect may not be significant. The exceptional cases, FRIs at high and FRIIs at low accretion rate, are exclusively at the high and low end of the distribution of the flux ratio of VLBA core to 178 MHz flux density. It is not impossible that the location of these sources are due to the recent shining or weakening of the central engine (i.e., both accretion and jet).

\section{ACKNOWLEDGEMENTS} \label{sec:rotate}

We thank the anonymous referee for constructive comments that improved the manuscript. We thank Minhua Zhou, Mai Liao,and Jiawen Li for helpful discussions. Special thanks are given to Robert Antonucci for the initialization of the project and valuable discussions. This work is supported by the National Science Foundation
of China (grants 11473054, U1531245, 11763002, and 11590784).
This research has made use of the NASA/IPAC Extragalactic Database (NED), which is operated by the Jet Propulsion Laboratory, California Institute of Technology, under contract with the National Aeronautics and Space Administration. The VLBA experiment is sponsored by Shanghai Astronomical Observatory through the MoU with the NRAO. The VLBA is operated by the Long Baseline Observatory which is managed by Associated
Universities, Inc., under cooperative agreement with the National
Science Foundation. The Long Baseline Observatory is a facility of the National Science Foundation operated under cooperative agreement by Associated Universities, Inc.

 \textit{Facility:} VLBA

{\it Software:} IDL, AIPS, DIFMAP

\newpage
\clearpage
\begin{figure}[h]
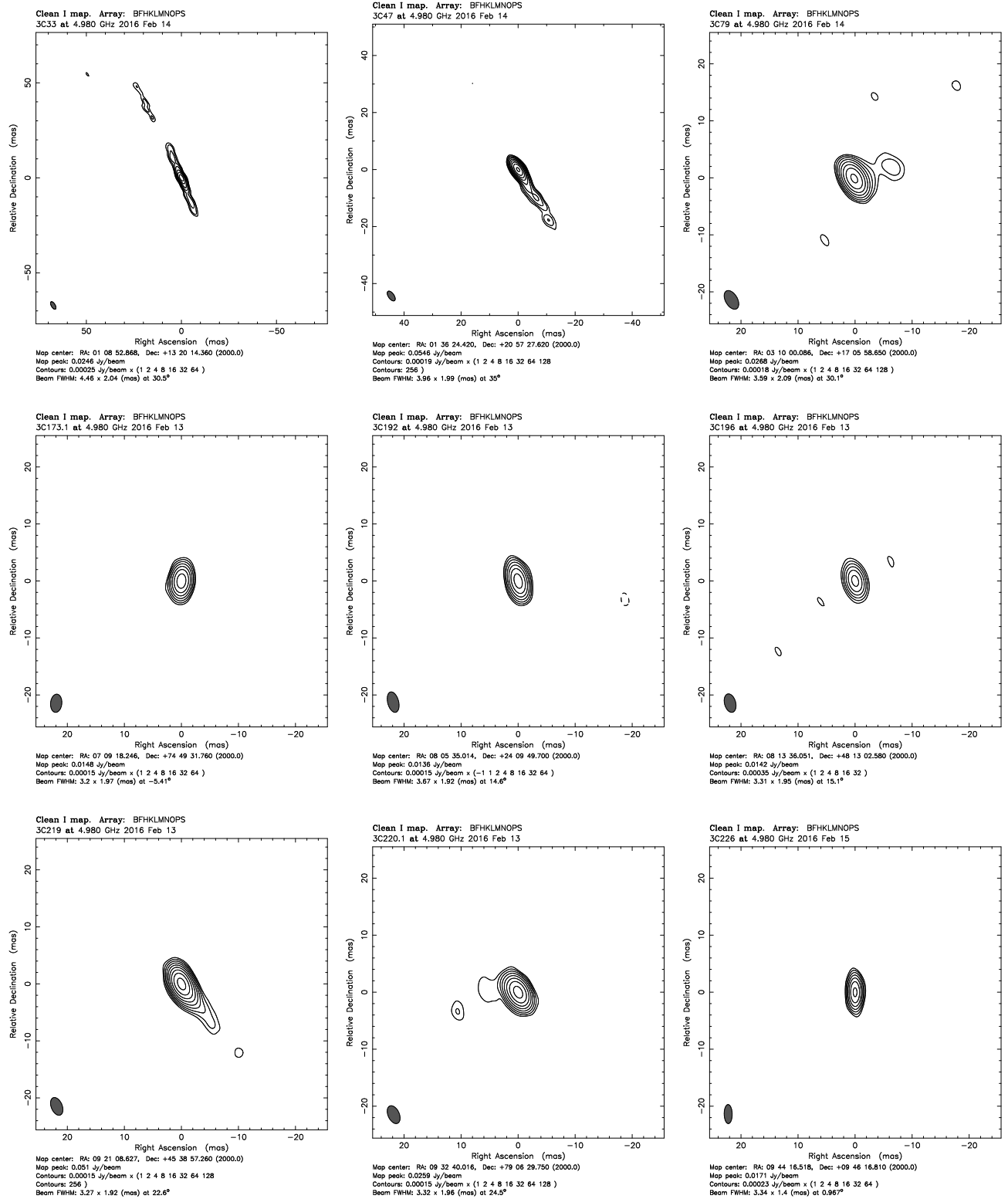

	\begin{minipage}[t]{0.4\linewidth}
		\centering
		\includegraphics[width=55mm]{3c33.eps}		
	\end{minipage}%
	\begin{minipage}[t]{0.4\textwidth}
		\centering
		\includegraphics[width=55mm]{3c47.eps}
	\end{minipage}%
	\begin{minipage}[t]{0.4\linewidth}
		\centering
		\includegraphics[width=55mm]{3c79.eps}
	\end{minipage}%
	
	\vspace*{0.5cm}
	
	\begin{minipage}[t]{0.4\linewidth}
		\centering
		\includegraphics[width=55mm]{3c173_1.eps}		
	\end{minipage}%
	\begin{minipage}[t]{0.4\textwidth}
		\centering
		\includegraphics[width=55mm]{3c192.eps}
	\end{minipage}%
	\begin{minipage}[t]{0.4\linewidth}
		\centering
		\includegraphics[width=55mm]{3c196.eps}
	\end{minipage}%
	
	\vspace*{0.5cm}
	
	\begin{minipage}[t]{0.4\linewidth}
		\centering
		\includegraphics[width=55mm]{3c219.eps}		
	\end{minipage}%
	\begin{minipage}[t]{0.4\textwidth}
		\centering
		\includegraphics[width=55mm]{3c220_1.eps}
	\end{minipage}%
	\begin{minipage}[t]{0.4\linewidth}
		\centering
		\includegraphics[width=55mm]{3c226.eps}
	\end{minipage}%
	\vspace*{0.5cm}
	\centering
	\caption{VLBA 5 GHz images for 21 sources from our observations.	
		\label{fig:1}}
\end{figure}
\addtocounter{figure}{-1}
\clearpage
\newpage
\begin{figure}[h]
	\begin{minipage}[t]{0.4\linewidth}
		\centering
		\includegraphics[width=55mm]{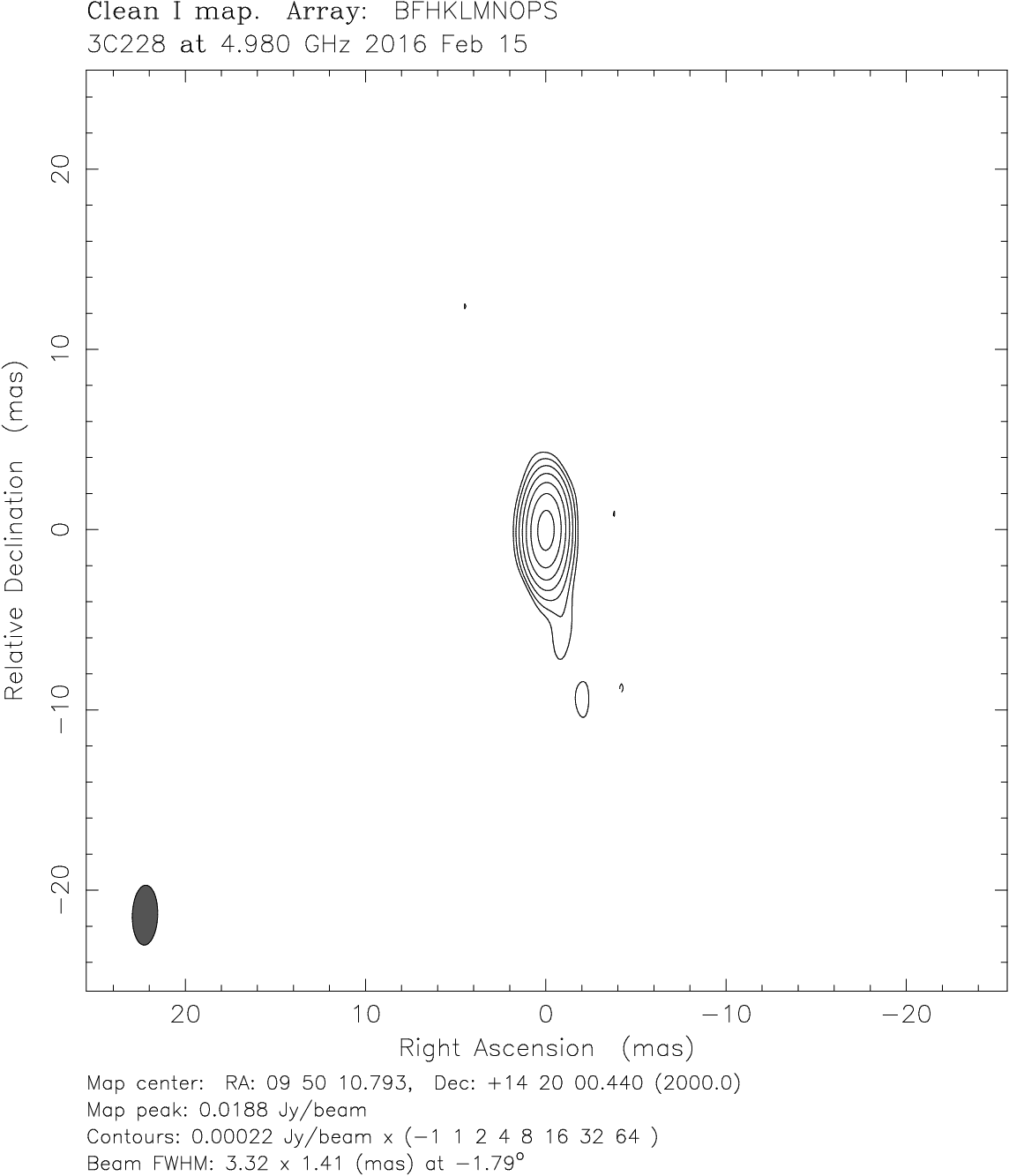}		
	\end{minipage}%
	\begin{minipage}[t]{0.4\textwidth}
		\centering
		\includegraphics[width=55mm]{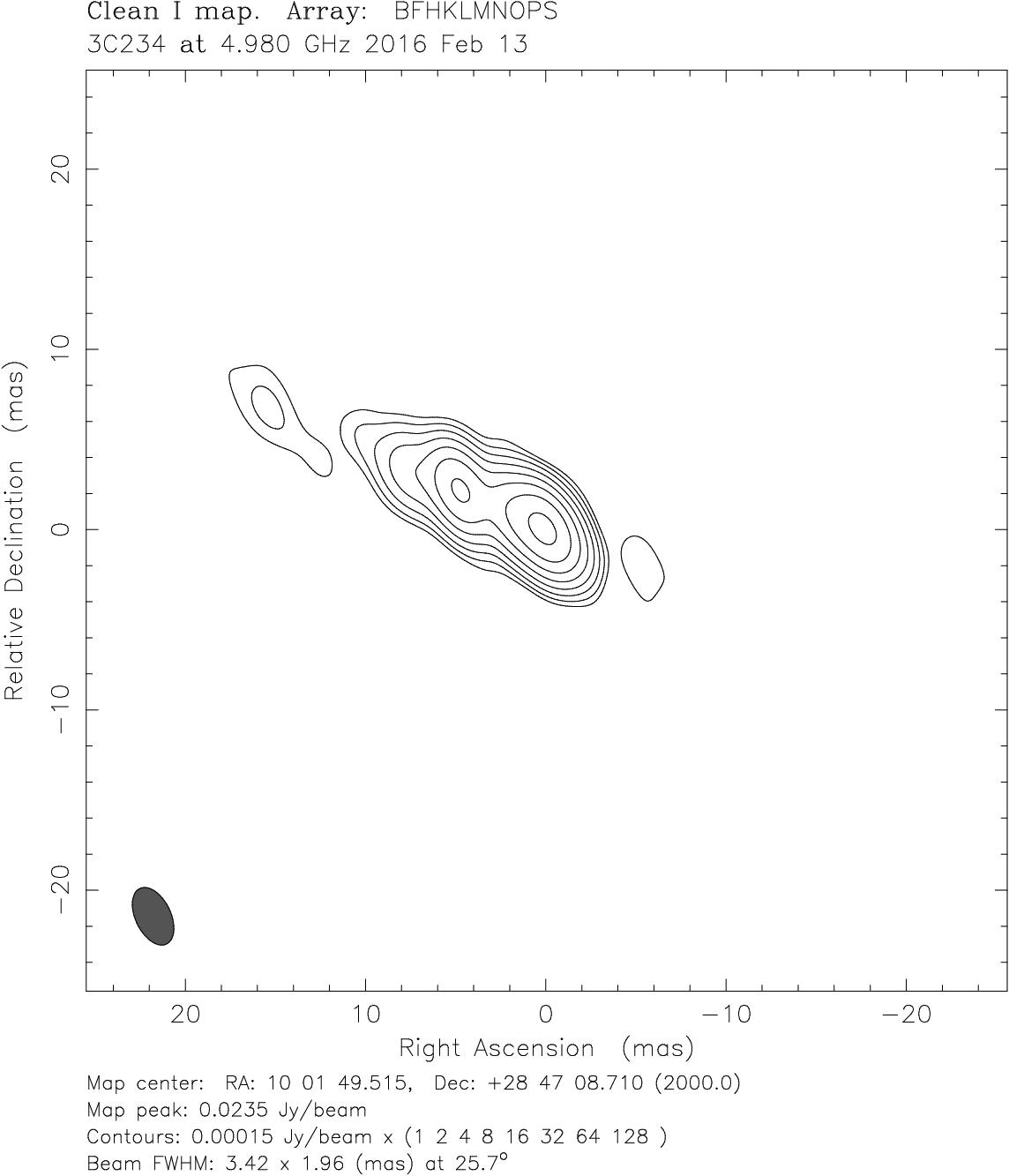}
	\end{minipage}%
	\begin{minipage}[t]{0.4\linewidth}
		\centering
		\includegraphics[width=55mm]{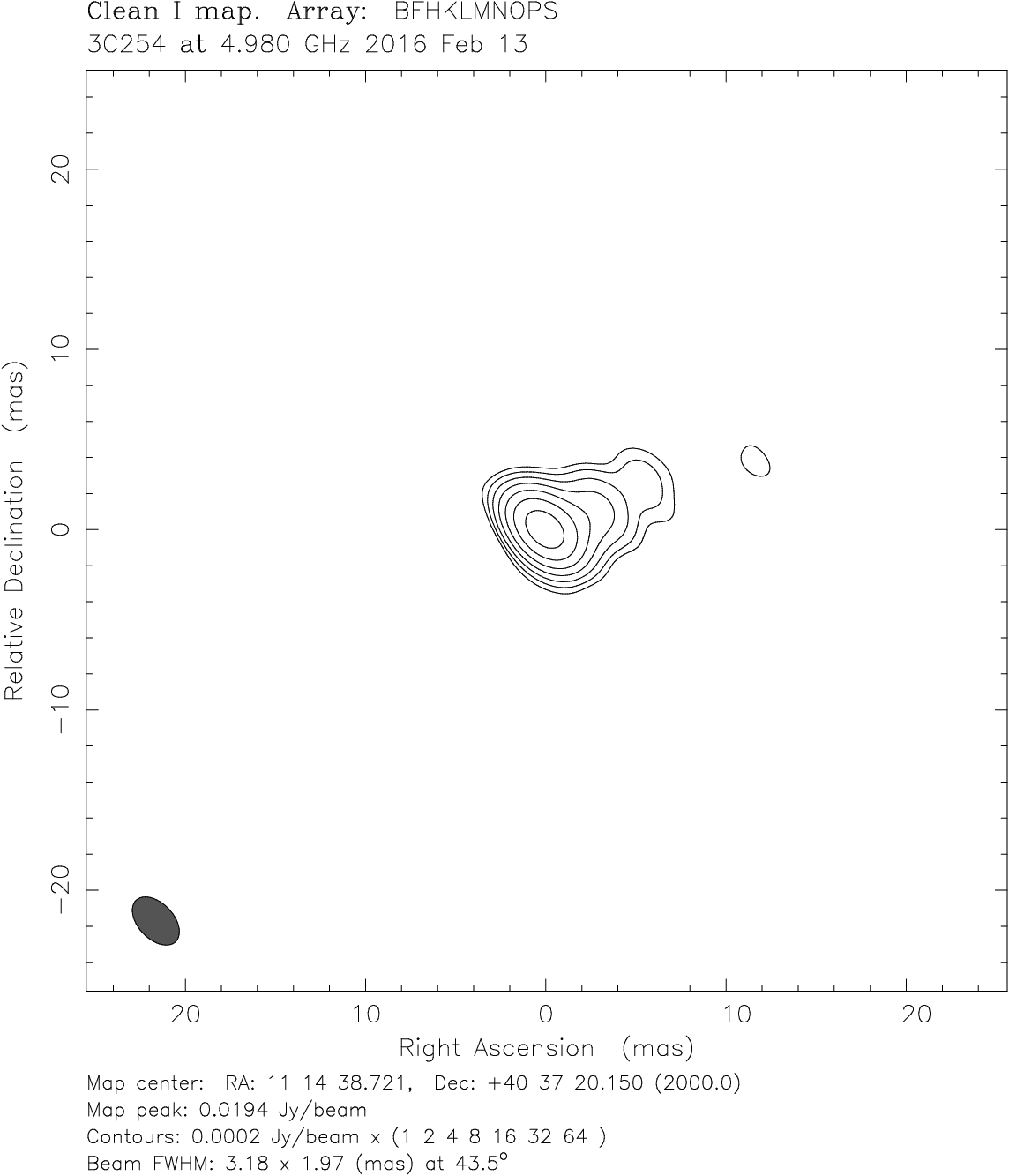}
	\end{minipage}%
	
	\vspace*{0.5cm}
	
	\begin{minipage}[t]{0.4\linewidth}
		\centering
		\includegraphics[width=55mm]{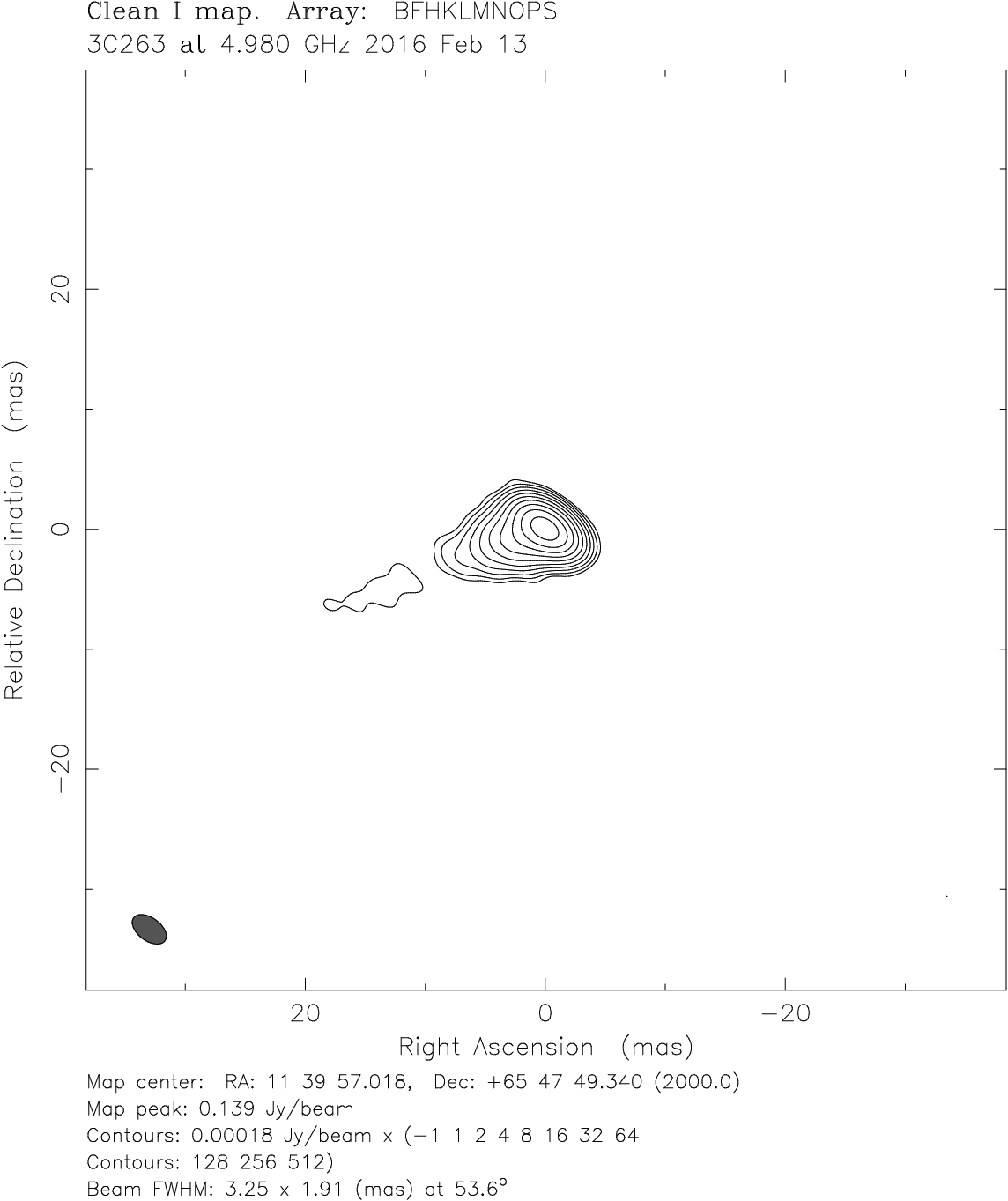}		
	\end{minipage}%
	\begin{minipage}[t]{0.4\textwidth}
		\centering
		\includegraphics[width=55mm]{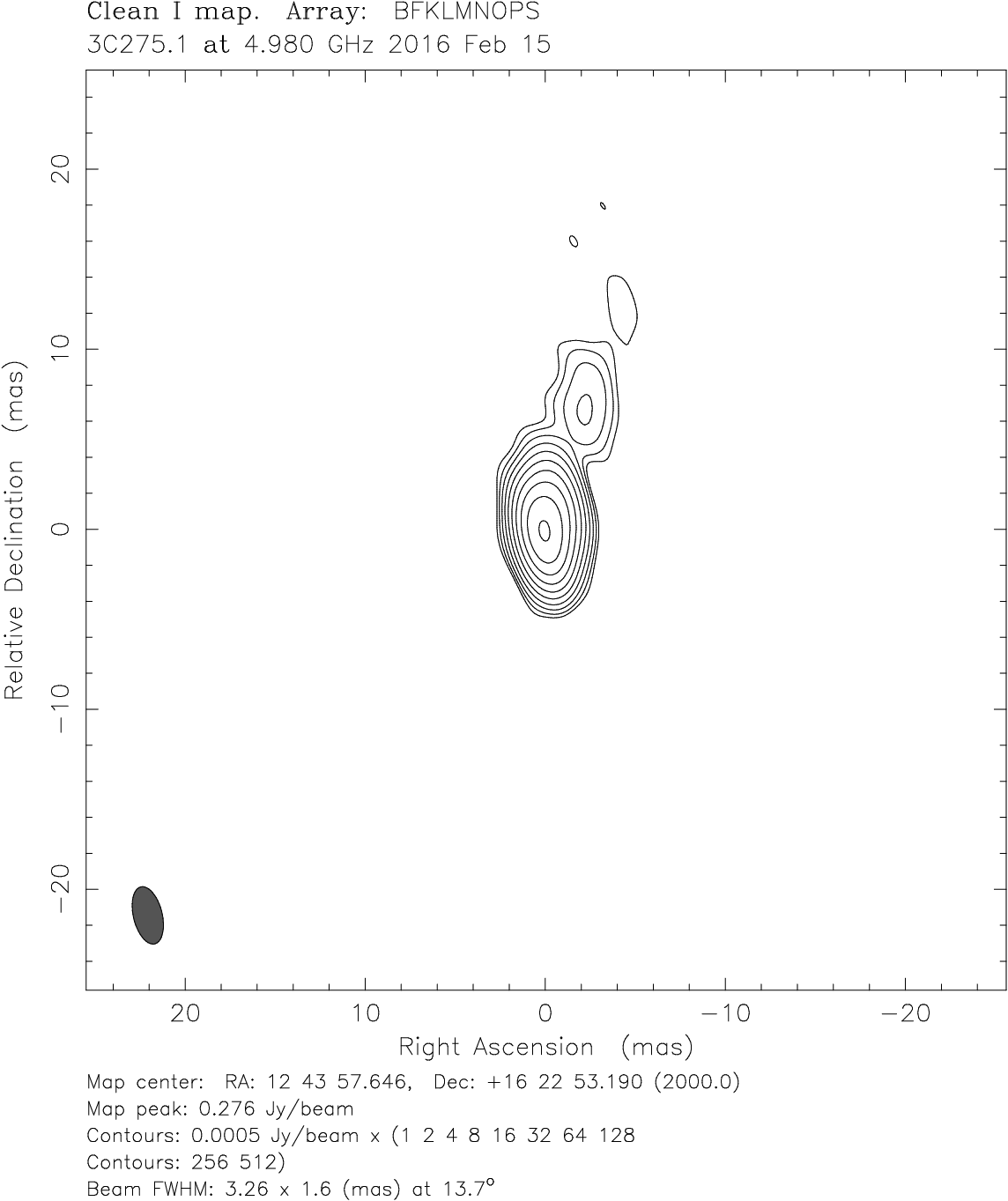}
	\end{minipage}%
	\begin{minipage}[t]{0.4\linewidth}
		\centering
		\includegraphics[width=55mm]{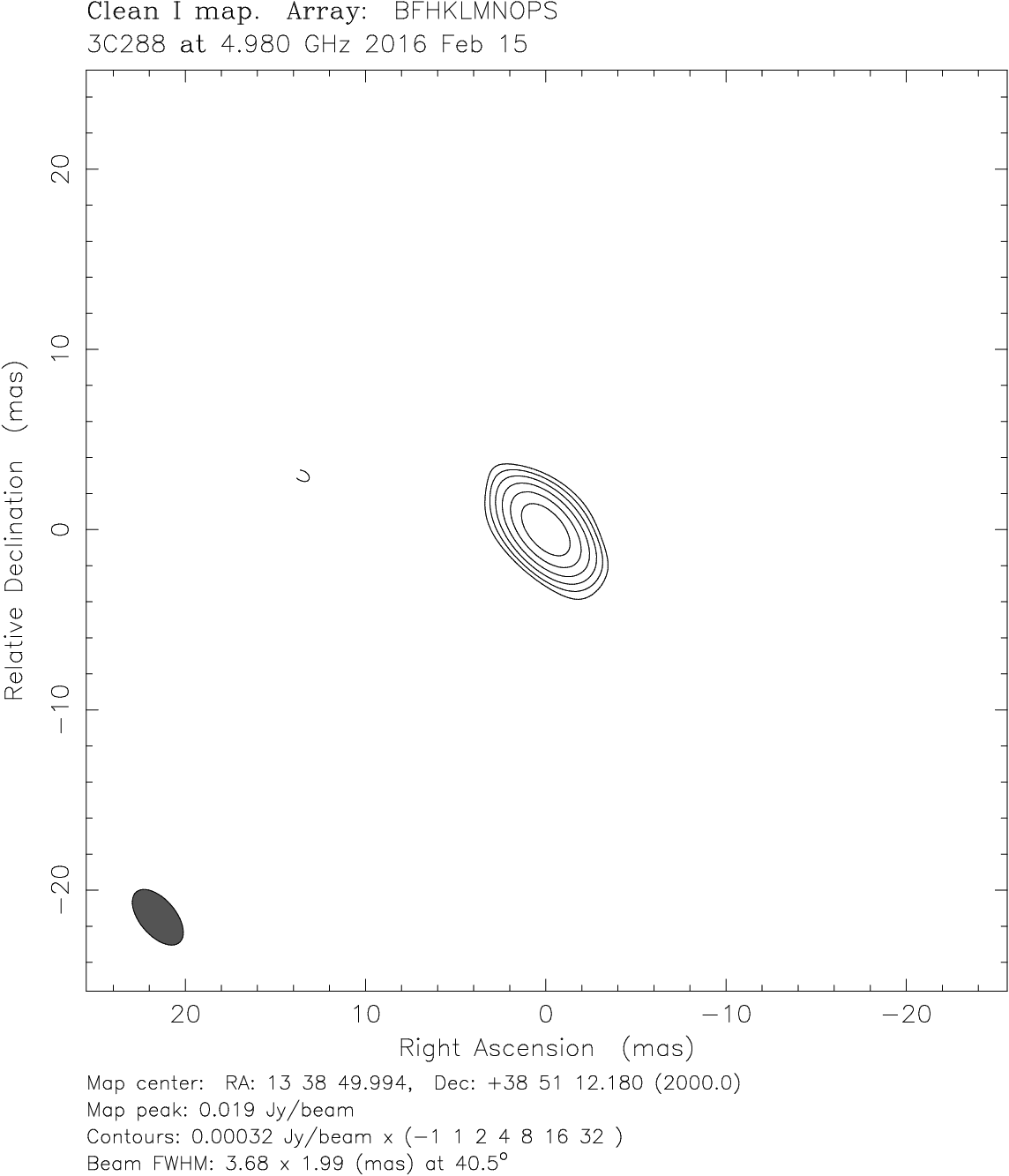}
	\end{minipage}%
	
	\vspace*{0.5cm}
	
	\begin{minipage}[t]{0.4\linewidth}
		\centering
		\includegraphics[width=55mm]{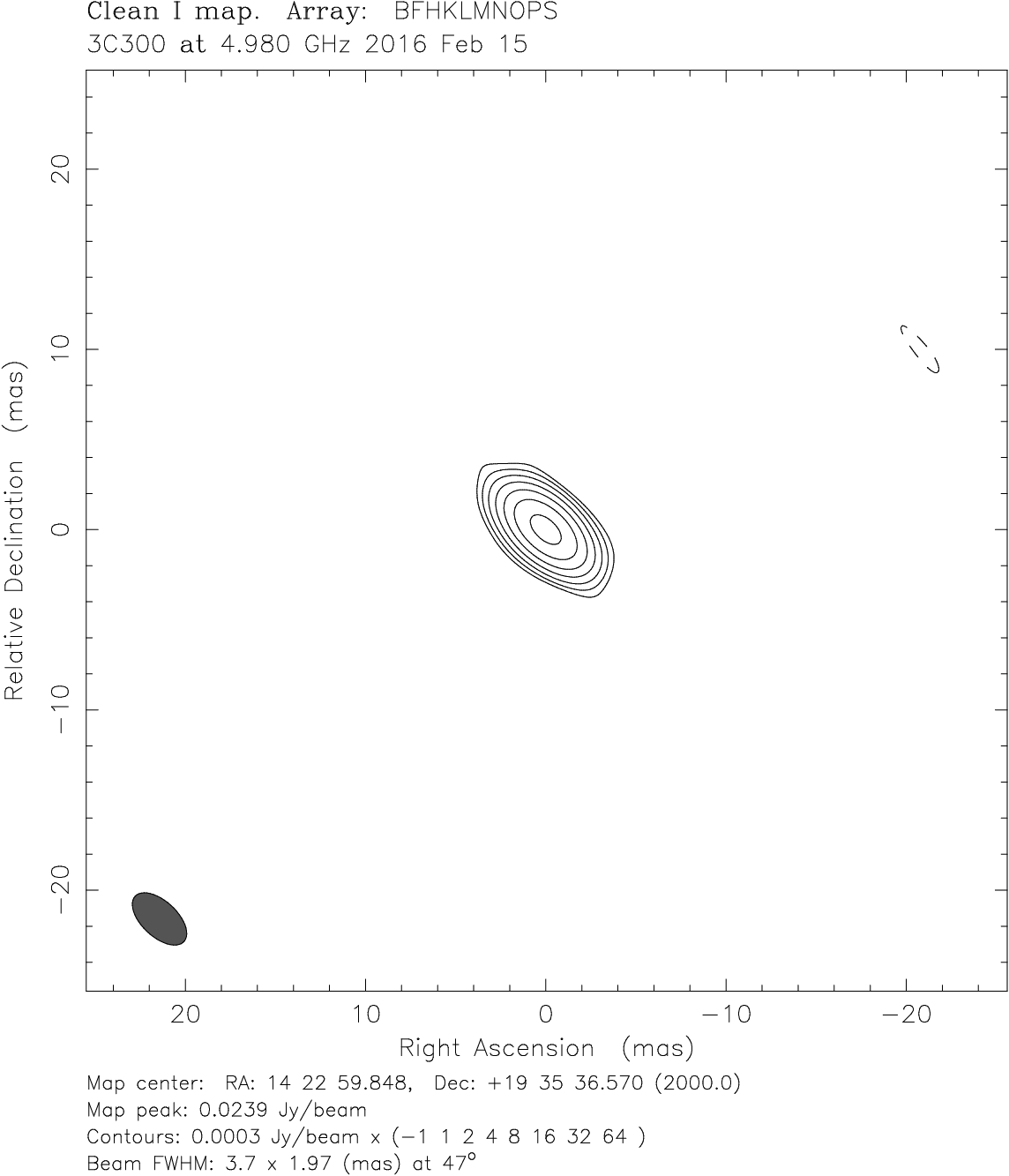}		
	\end{minipage}%
	\begin{minipage}[t]{0.4\textwidth}
		\centering
		\includegraphics[width=55mm]{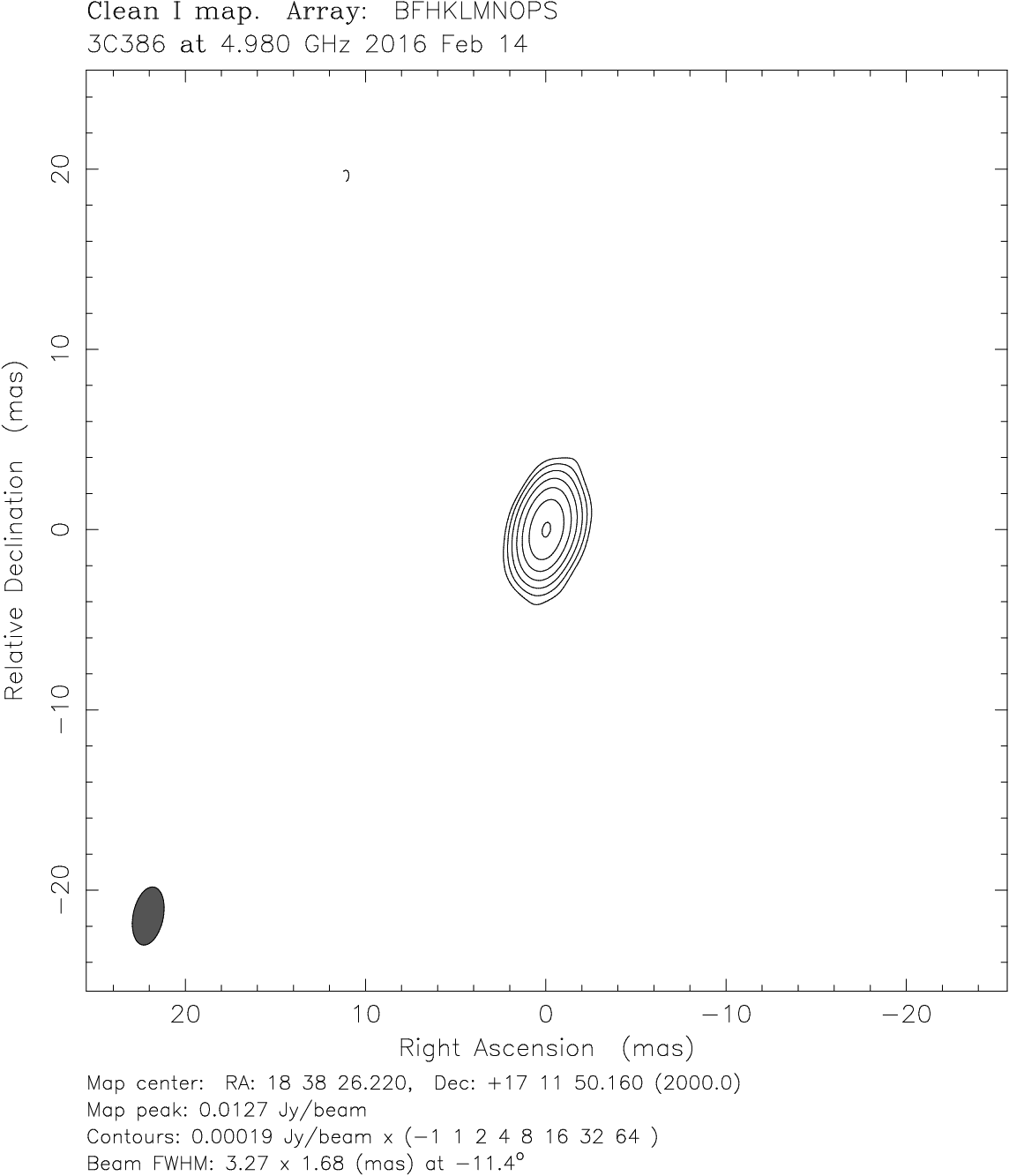}
	\end{minipage}%
	\begin{minipage}[t]{0.4\linewidth}
		\centering
		\includegraphics[width=55mm]{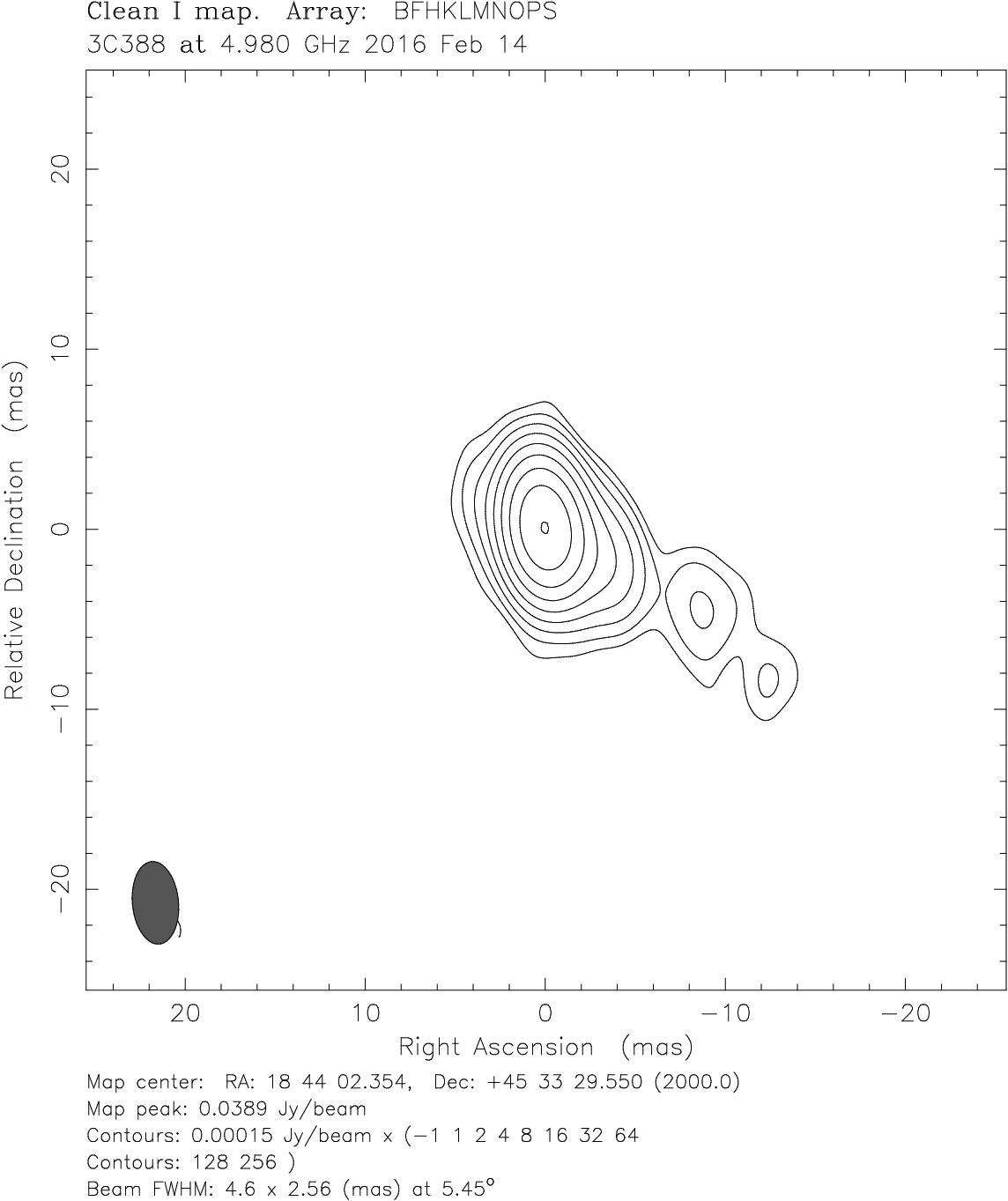}
	\end{minipage}%
	\caption{$-$continued.
		\label{fig:1}}	
\end{figure}
\addtocounter{figure}{-1}

\clearpage
\newpage
\vspace*{0.5cm}
\begin{figure}[h]
	\begin{minipage}[t]{0.4\linewidth}
		\centering
		\includegraphics[width=55mm]{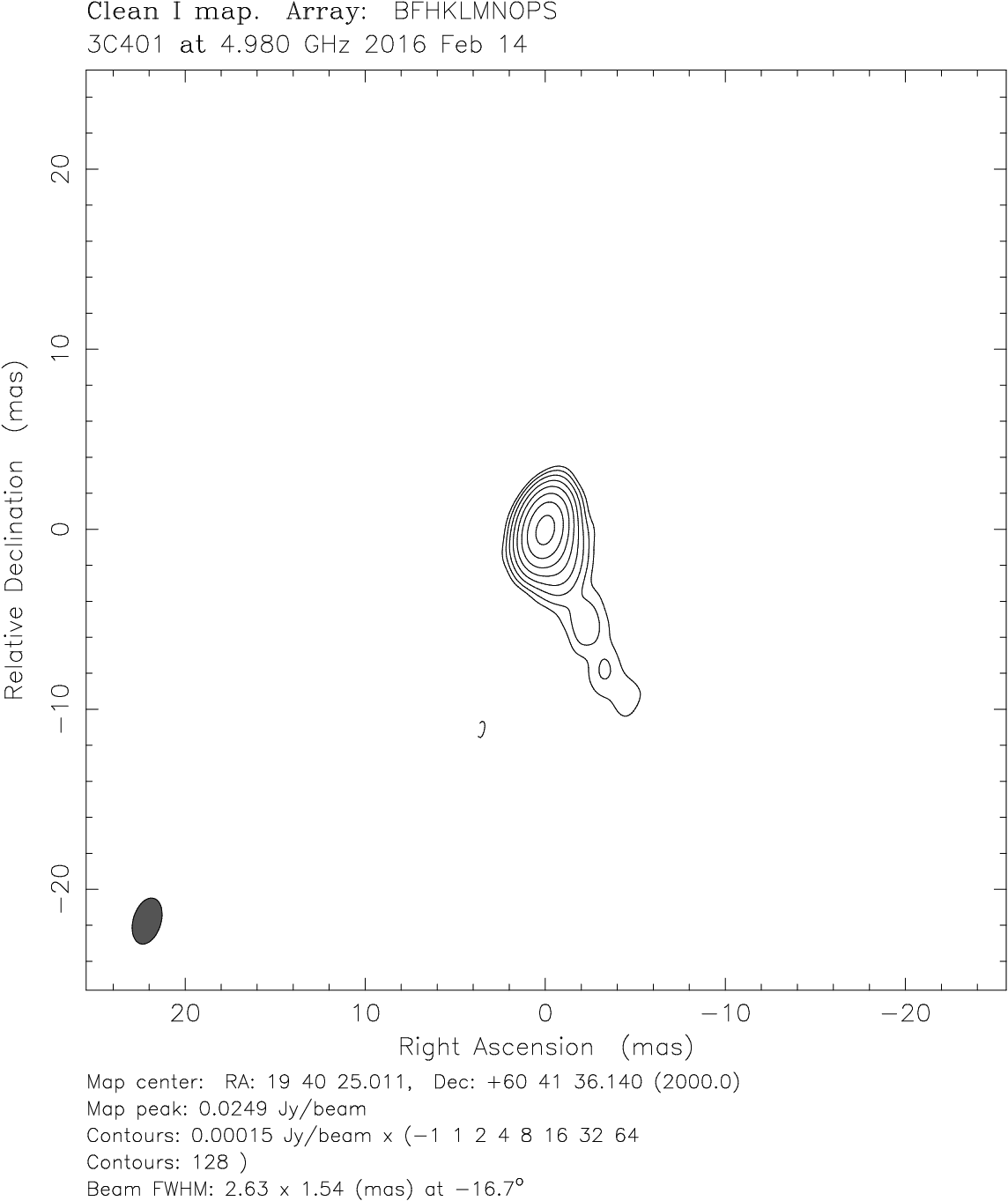}		
	\end{minipage}%
	\begin{minipage}[t]{0.4\textwidth}
		\centering
		\includegraphics[width=55mm]{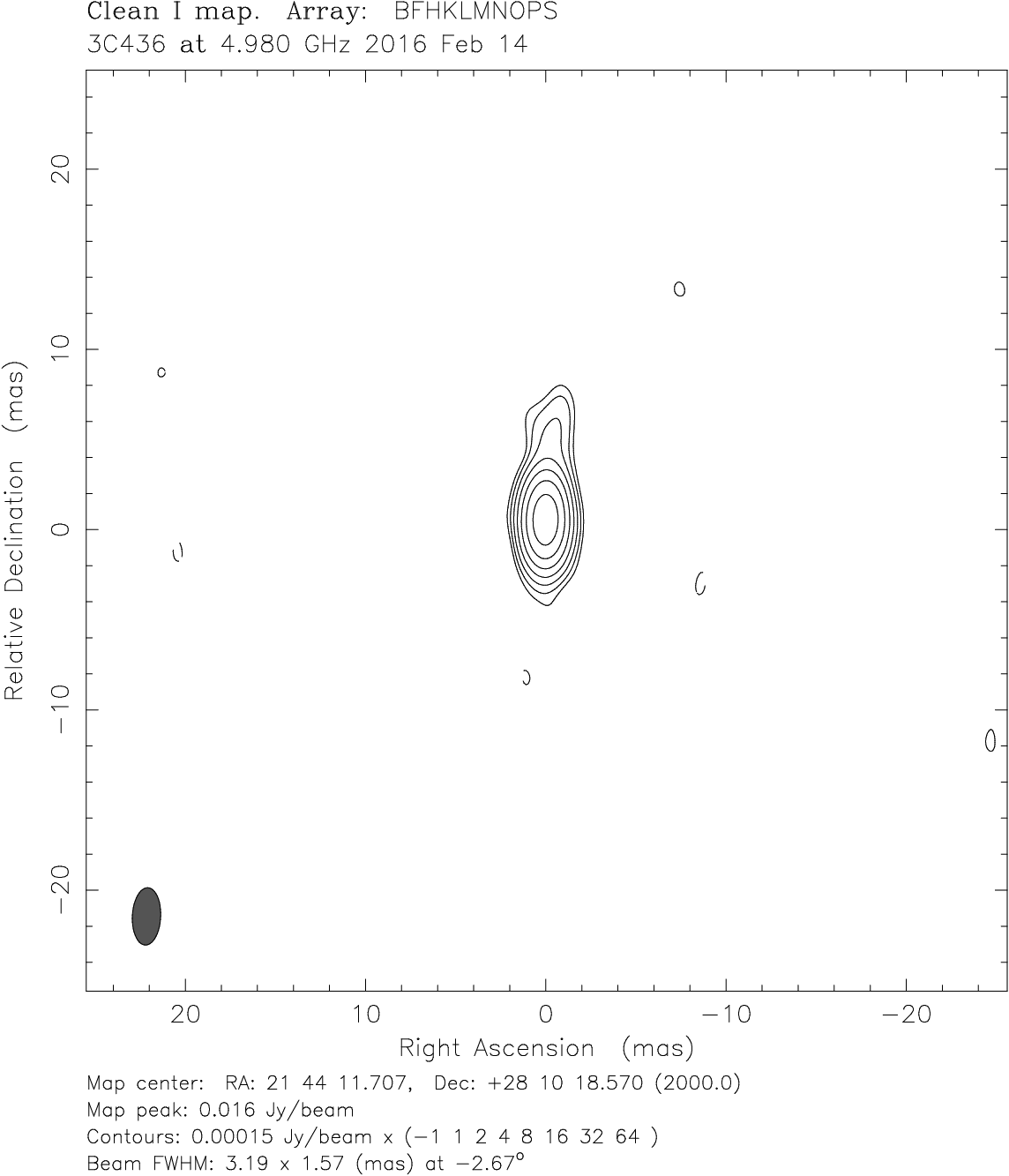}
	\end{minipage}%
	\begin{minipage}[t]{0.4\linewidth}
		\centering
		\includegraphics[width=55mm]{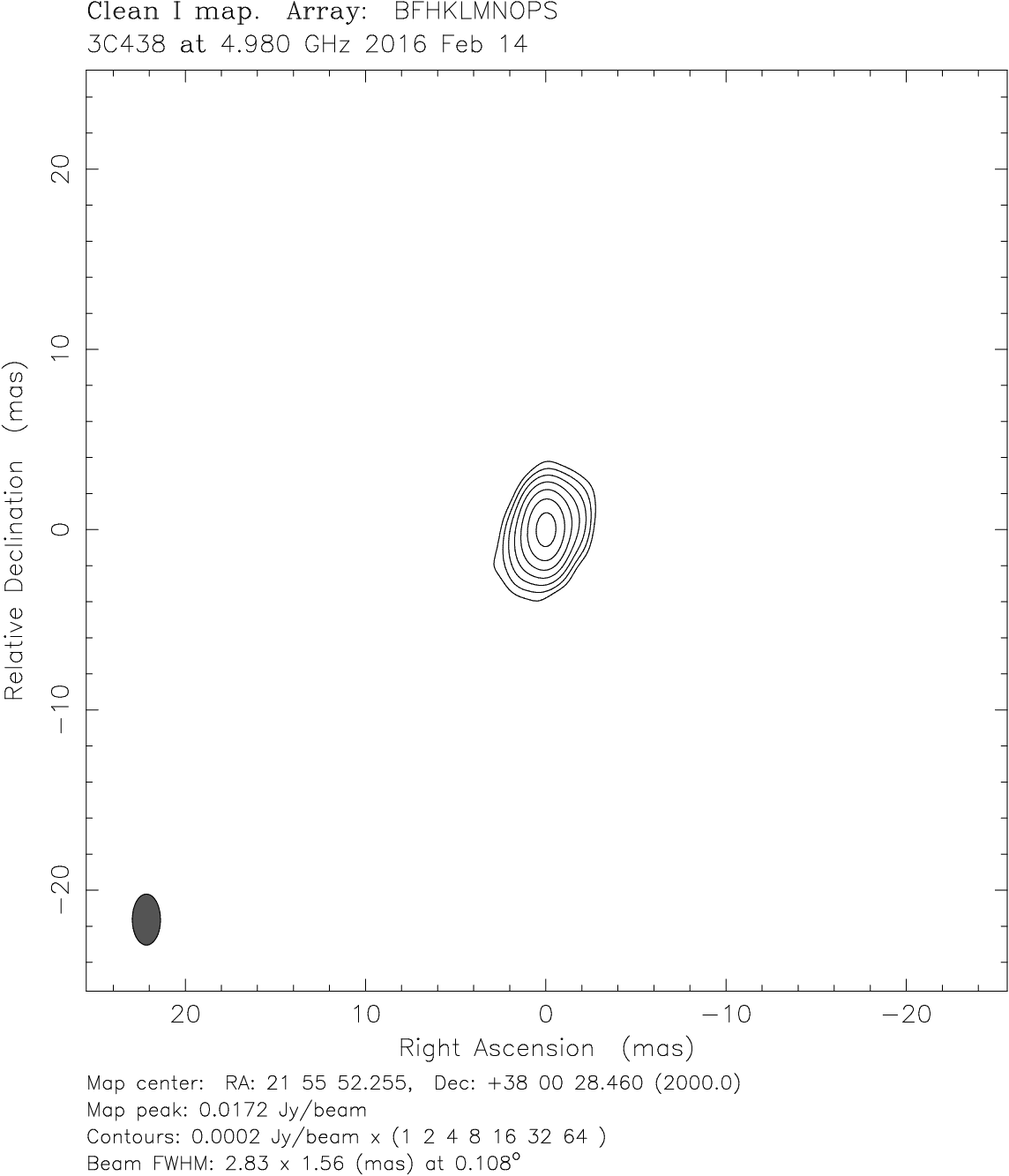}
	\end{minipage}%
	\caption{$-$ continued. \label{fig:bg239}}
\end{figure}
\newpage
\clearpage
\begin{figure}[h]
	\begin{minipage}[t]{0.4\linewidth}
		\centering
		\includegraphics[width=53mm]{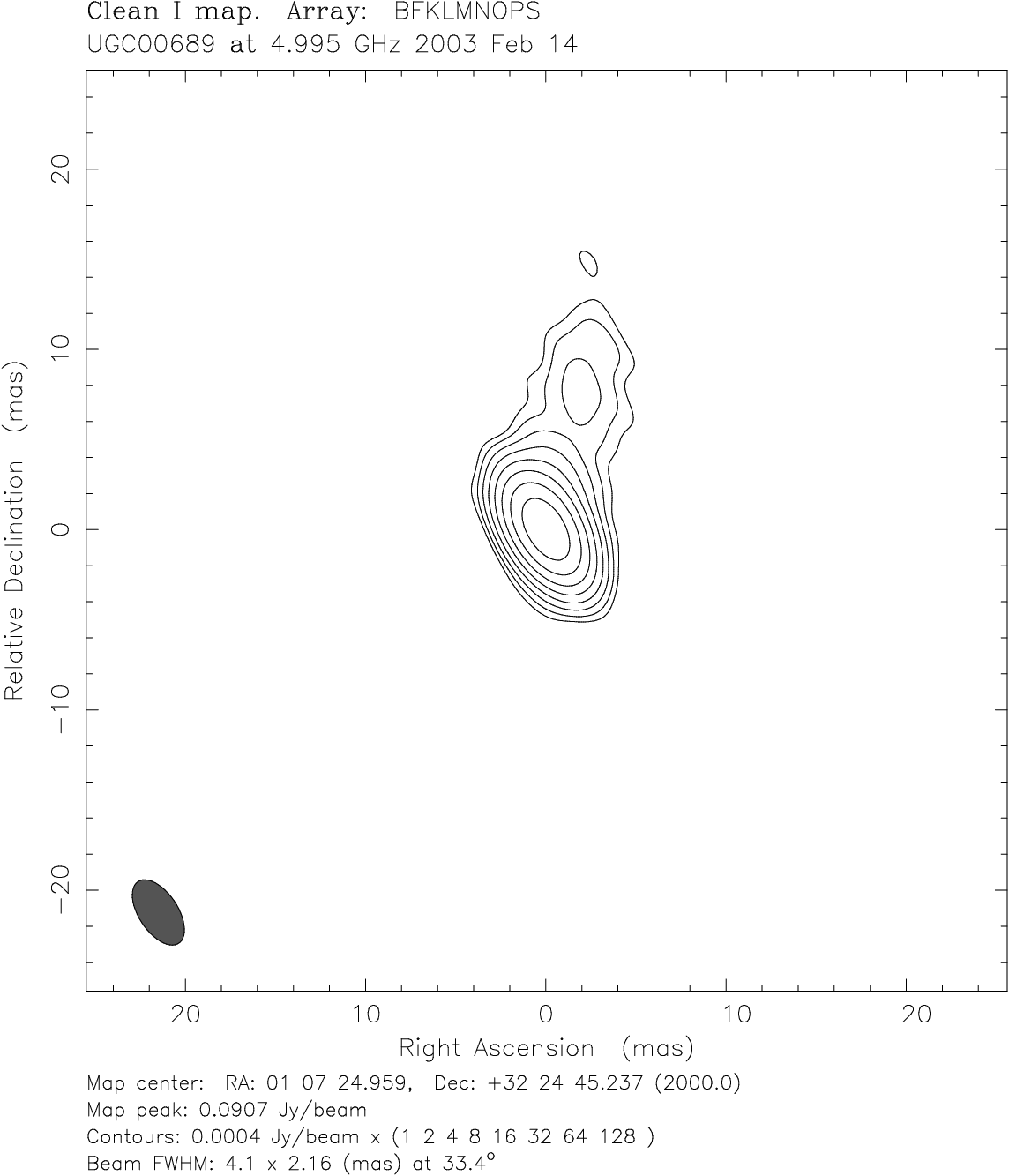}		
	\end{minipage}%
	\begin{minipage}[t]{0.4\textwidth}
		\centering
		\includegraphics[width=53mm]{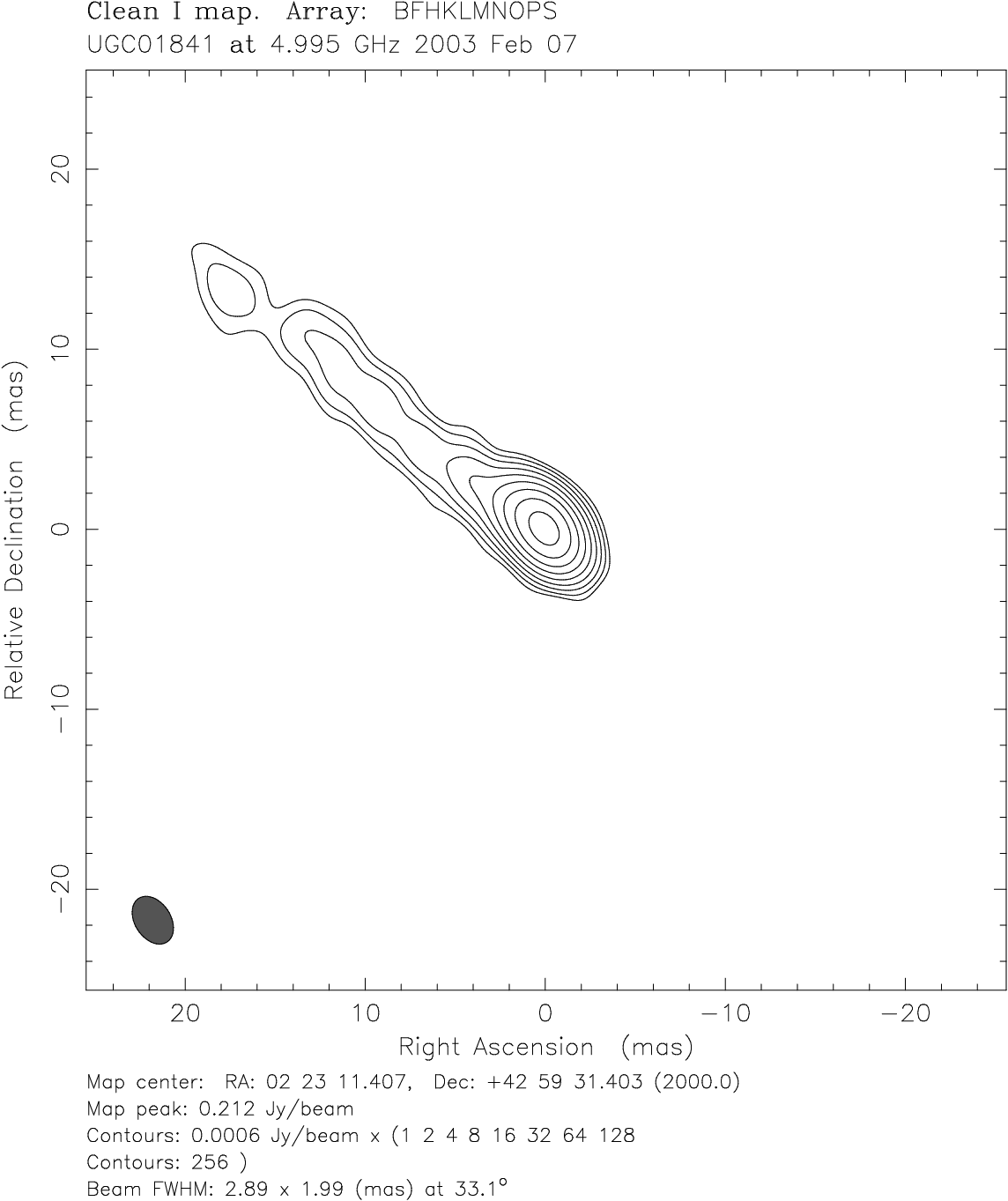}
	\end{minipage}%
	\begin{minipage}[t]{0.4\linewidth}
		\centering
		\includegraphics[width=53mm]{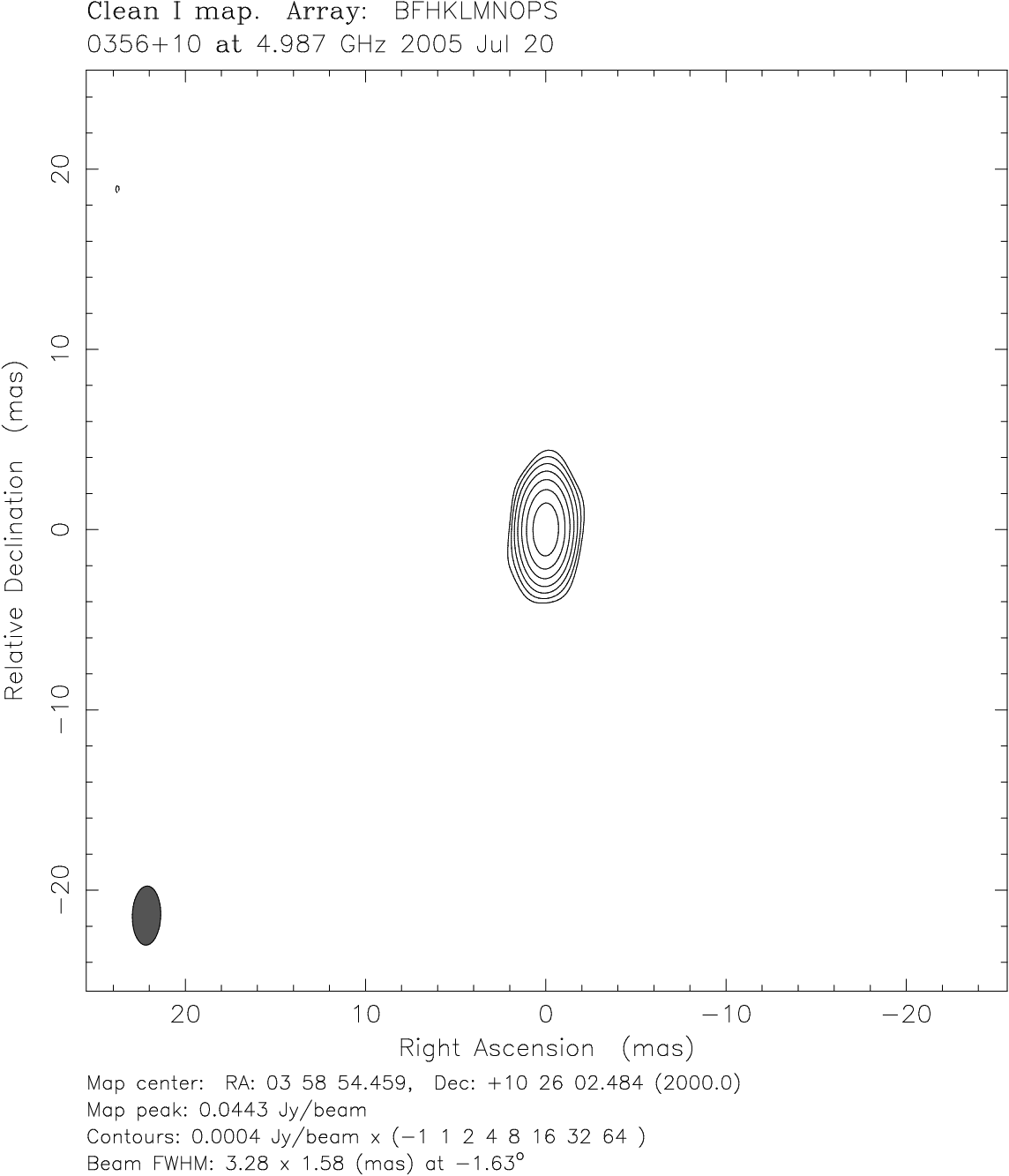}
	\end{minipage}%
	
	\begin{minipage}[t]{0.4\linewidth}
		\centering
		\includegraphics[width=53mm]{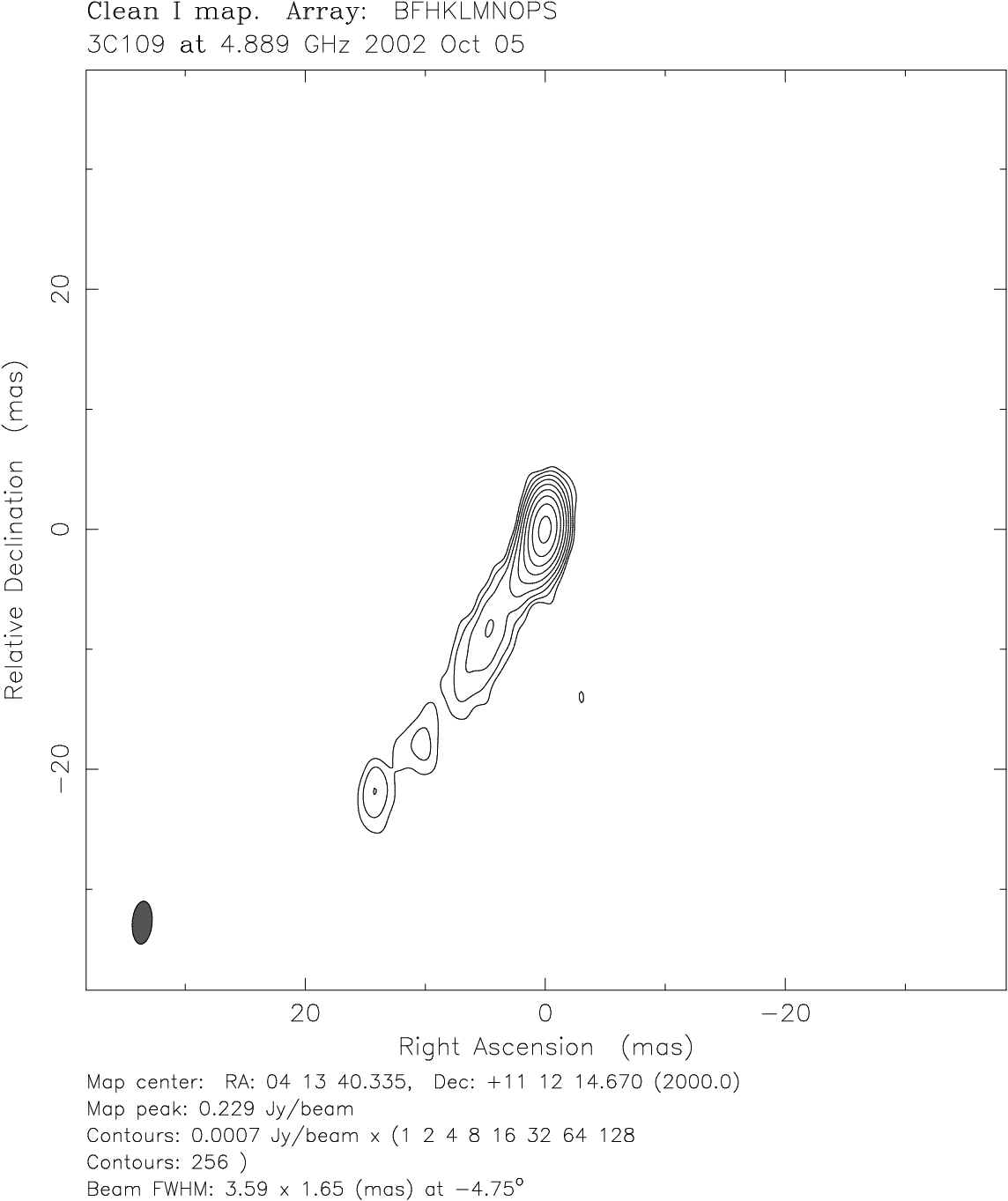}		
	\end{minipage}%
	\begin{minipage}[t]{0.4\textwidth}
		\centering
		\includegraphics[width=53mm]{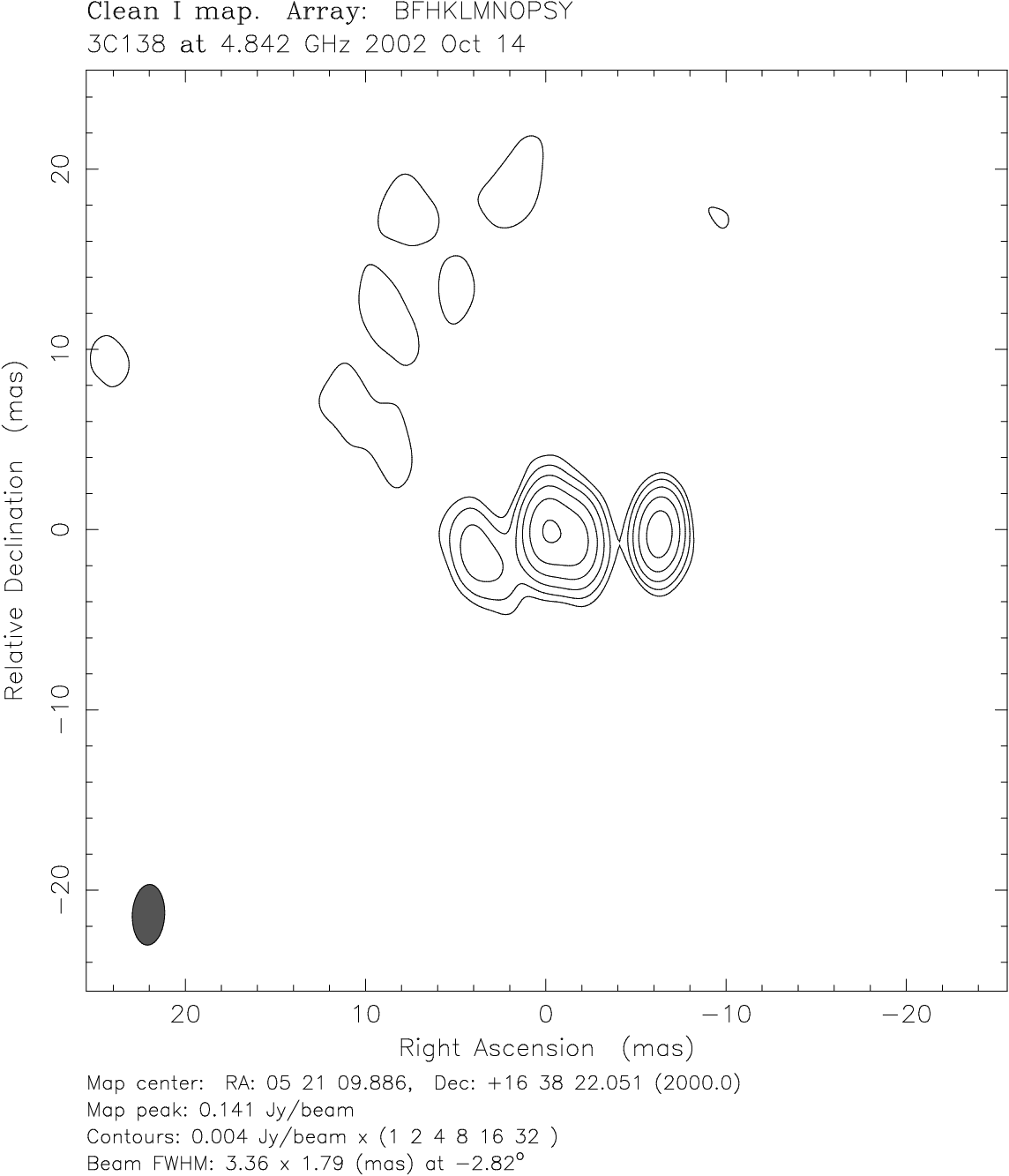}
	\end{minipage}%
	\begin{minipage}[t]{0.4\linewidth}
		\centering
		\includegraphics[width=53mm]{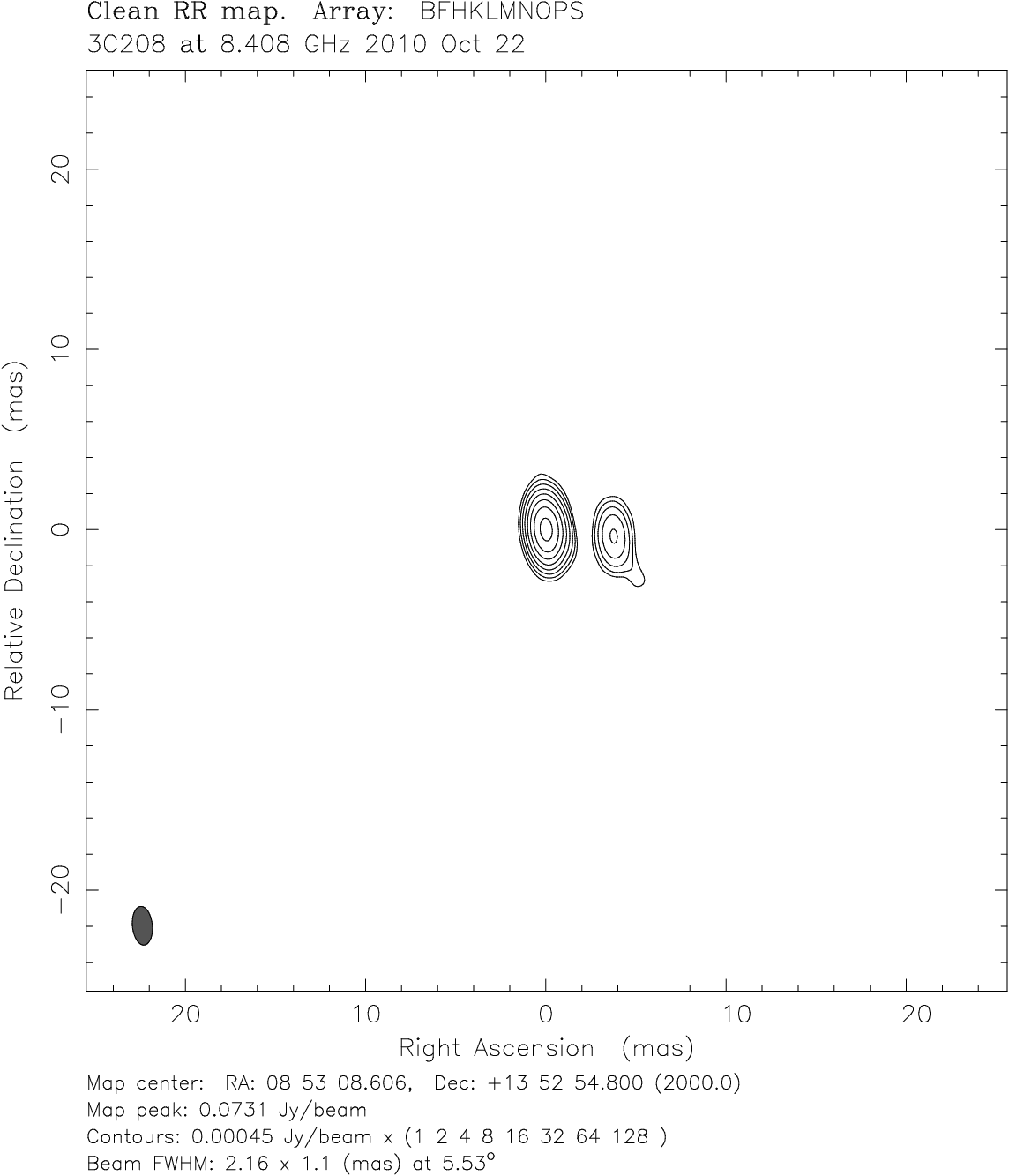}
	\end{minipage}%
	
	\vspace*{0.5cm}
	
	\begin{minipage}[t]{0.4\linewidth}
		\centering
		\includegraphics[width=53mm]{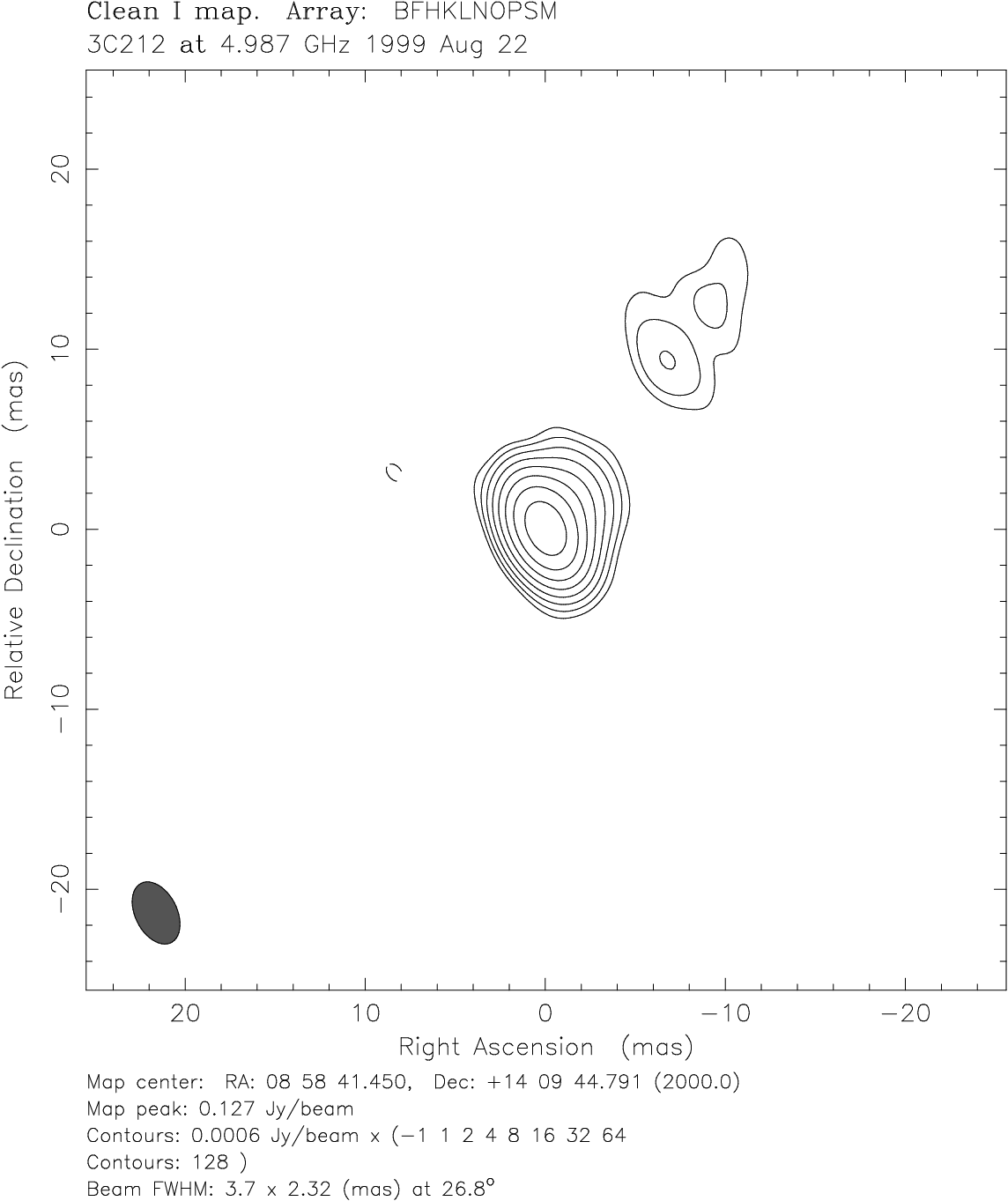}		
	\end{minipage}%
	\begin{minipage}[t]{0.4\textwidth}
		\centering
		\includegraphics[width=53mm]{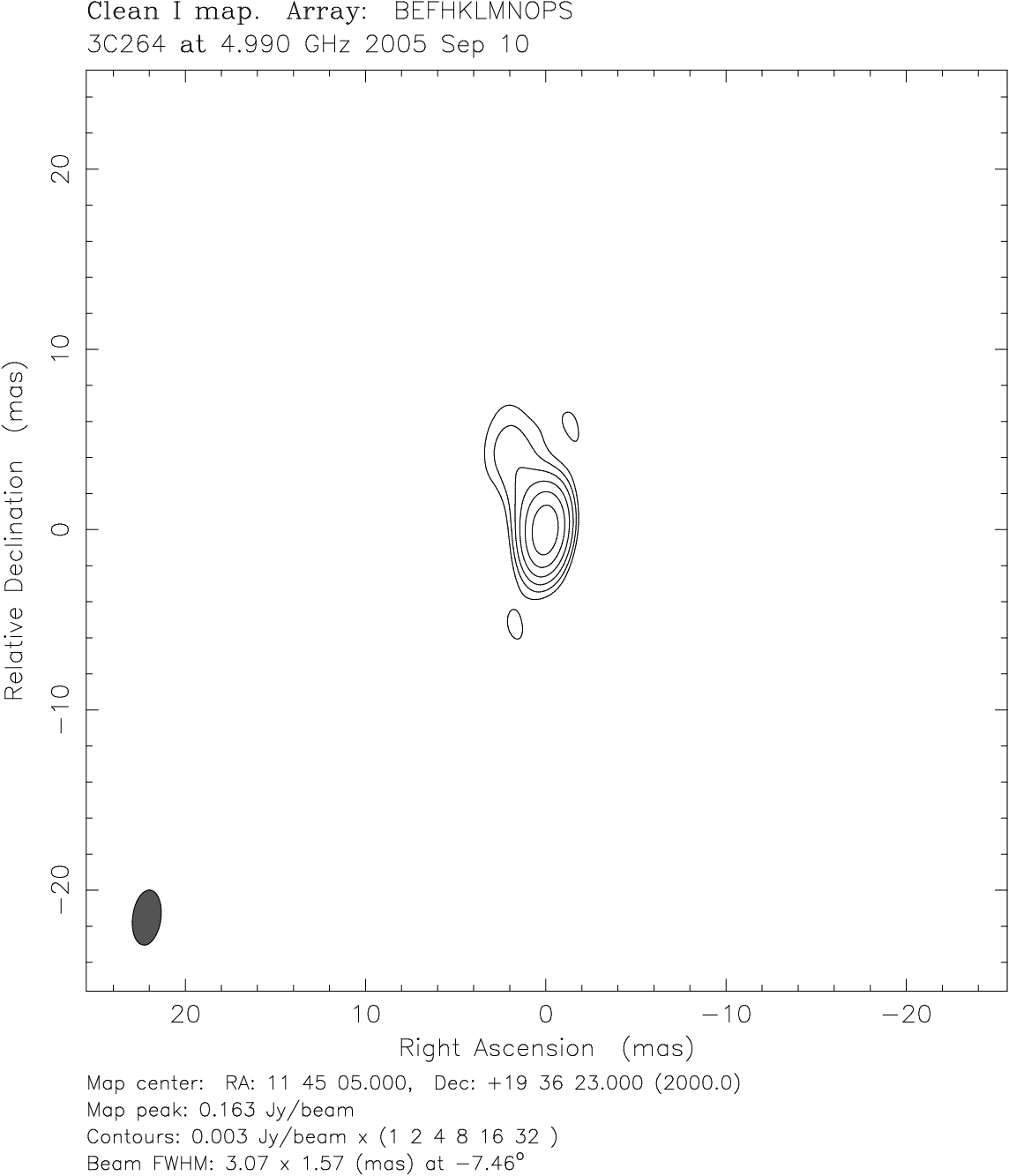}
	\end{minipage}%
	\begin{minipage}[t]{0.4\linewidth}
		\centering
		\includegraphics[width=53mm]{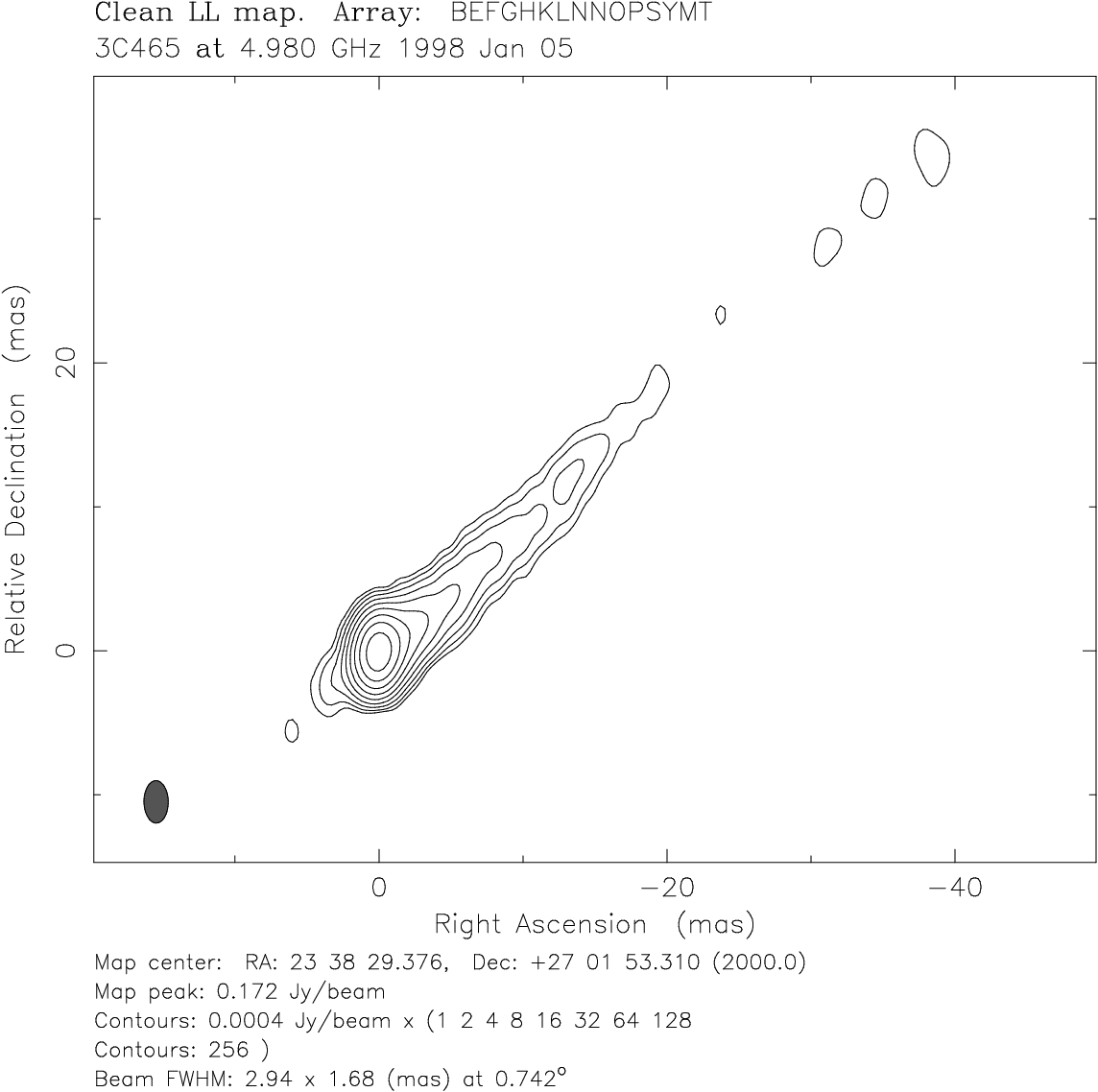}
	\end{minipage}%
	\vspace*{0.5cm}
	\centering
	\caption{VLBA 5 GHz images for 15 sources, and 8 GHz image for 3C 208, analyzed from NARO archival data.
		\label{fig:2}}
\end{figure}
\addtocounter{figure}{-1}
\clearpage
\newpage
\begin{figure}[h]
	\begin{minipage}[t]{0.4\linewidth}
		\centering
		\includegraphics[width=55mm]{3c309_1.eps}		
	\end{minipage}%
	\begin{minipage}[t]{0.4\textwidth}
		\centering
		\includegraphics[width=55mm]{3c326.eps}
	\end{minipage}%
	\begin{minipage}[t]{0.4\linewidth}
		\centering
		\includegraphics[width=55mm]{3c338.eps}
	\end{minipage}%
	
	
	\begin{minipage}[t]{0.4\linewidth}
		\centering
		\includegraphics[width=55mm]{3c380.eps}		
	\end{minipage}%
	\begin{minipage}[t]{0.4\textwidth}
		\centering
		\includegraphics[width=55mm]{3c382.eps}
	\end{minipage}%
	\begin{minipage}[t]{0.4\linewidth}
		\centering
		\includegraphics[width=55mm]{3c452.eps}
	\end{minipage}%

	\vspace*{0.5cm}
	\begin{minipage}[t]{0.4\linewidth}
		\centering		
		\includegraphics[width=55mm]{3c274.eps}	
	\end{minipage}
	\caption{$-$ continued. \label{fig:other}}
\end{figure}
\begin{figure*}
	\begin{center}
		\includegraphics[height=0.35\textheight]{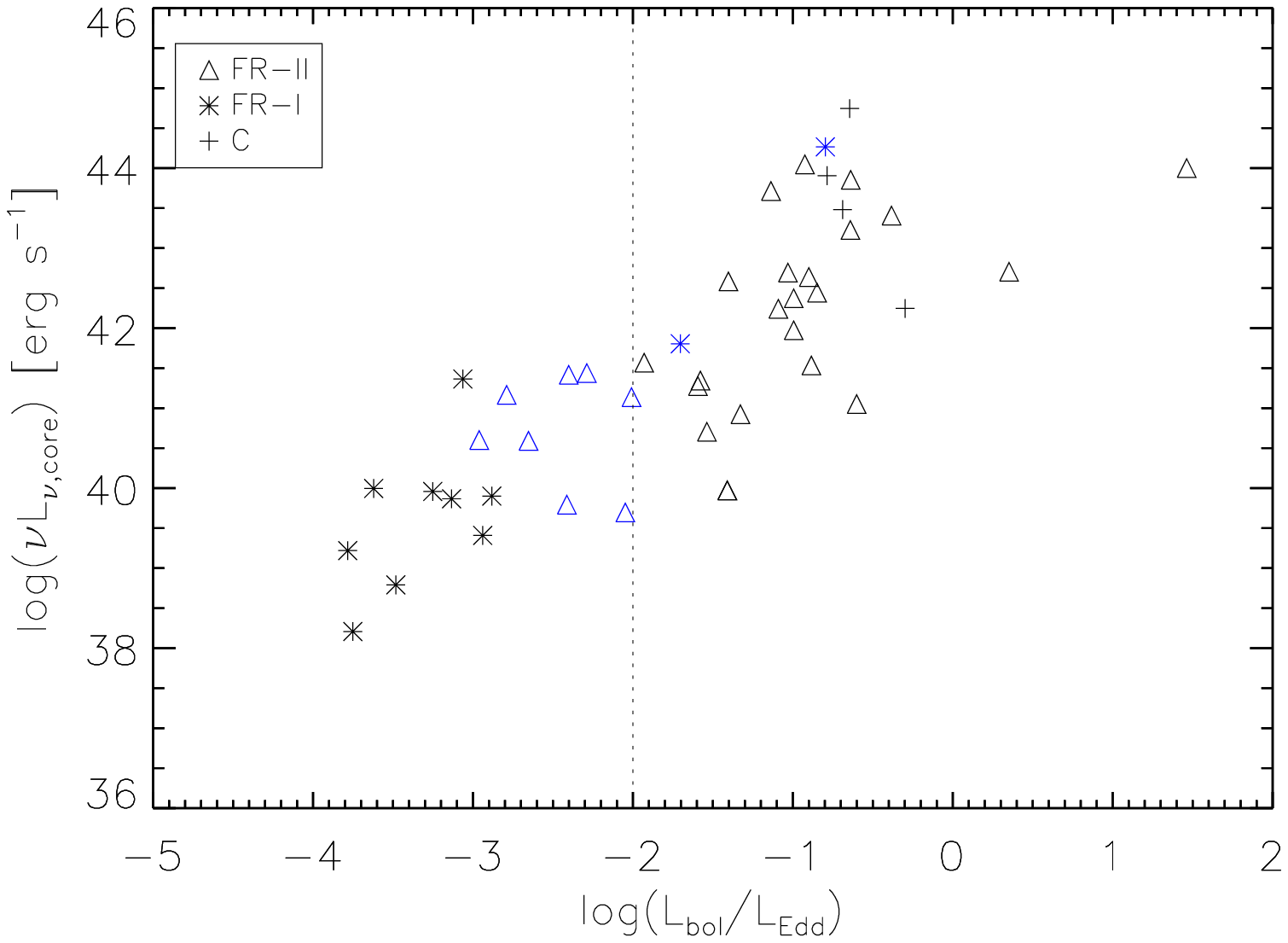}
	\end{center}	
	
	\caption{ The VLBA core luminosity at 5 GHz versus the Eddington ratio. The asterisks are for FRIs, the triangles for FRIIs, and crosses represent core-dominated sources. The Eddington ratio $L_{\rm bol}/L_{\rm Edd}=0.01$ is shown as the dotted line to distinguish the high and low accretion rate.\label{fig:F_vlba}}
\end{figure*}

\begin{figure*}
	\begin{center}
		\includegraphics[height=0.35\textheight]{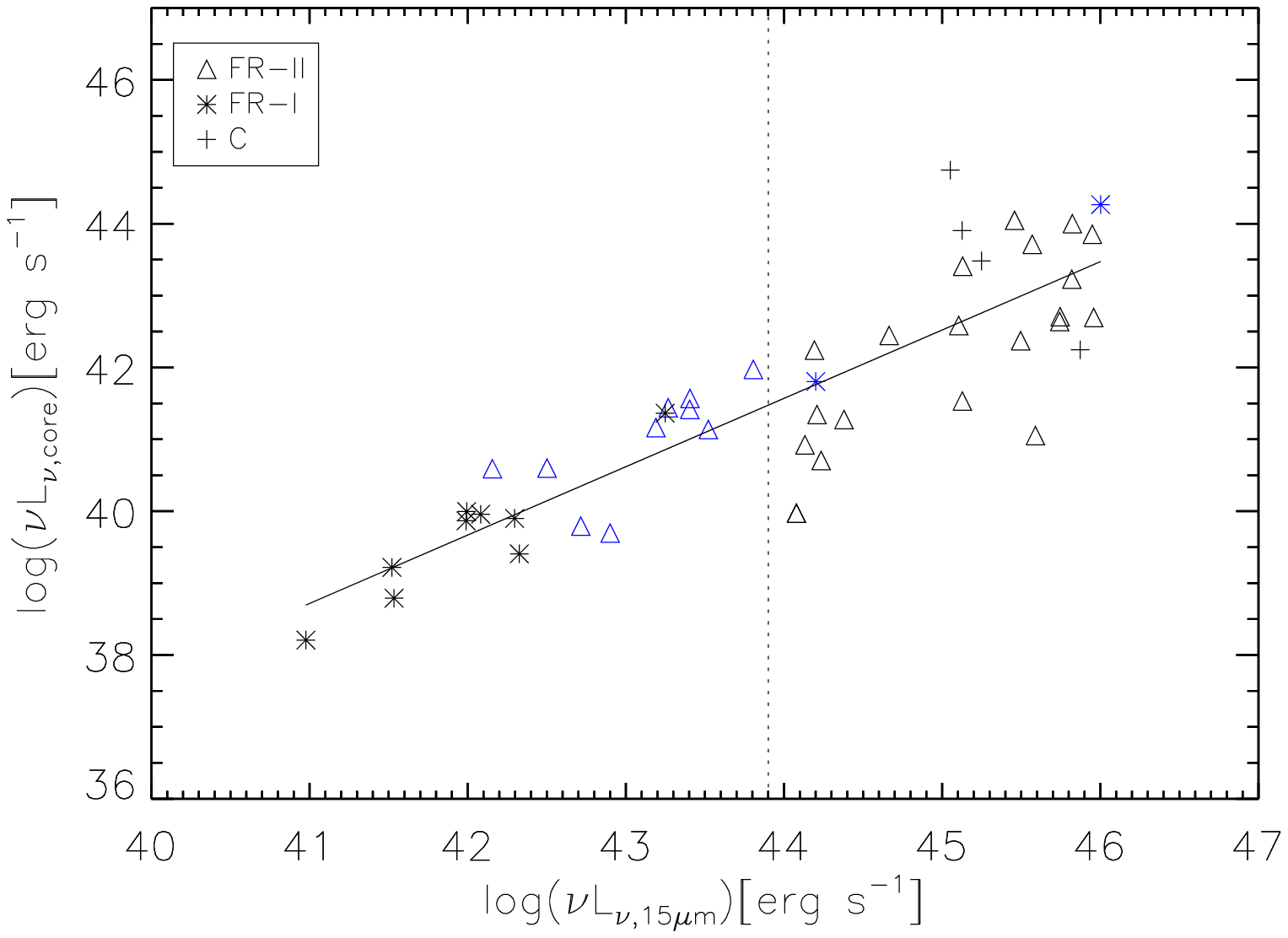}
	\end{center}	
	
	\caption{The VLBA core luminosity at 5 GHz versus the MIR luminosity at 15 $\rm \mu m$. The solid line is the linear fit. The dotted line is $\nu L_{\nu,\rm 15 \mu m}=8\times10^{43} \rm erg ~s^{-1}$, used in Ogle et al. (\cite{2006ApJ...647..161O}) to distinguish the MIR-weak and MIR-luminous radio galaxies.  
		\label{fig:MIR_core}}
\end{figure*}

\newpage
\clearpage
\begin{figure}[h]
	\begin{minipage}[t]{0.5\linewidth}
		\centering
		\includegraphics[width=70mm]{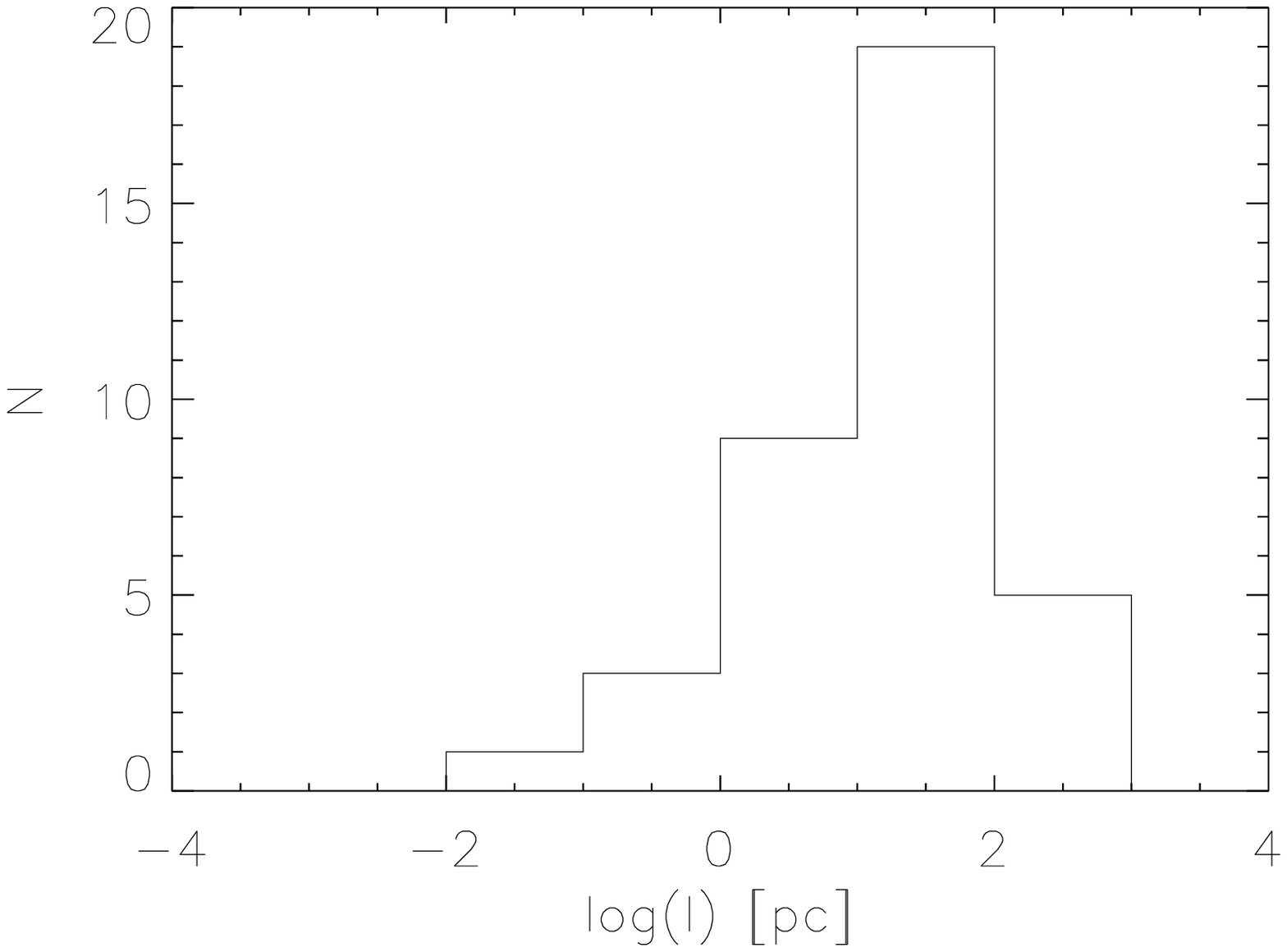}		
	\end{minipage}%
	\begin{minipage}[t]{0.5\textwidth}
		\centering
		\includegraphics[width=70mm]{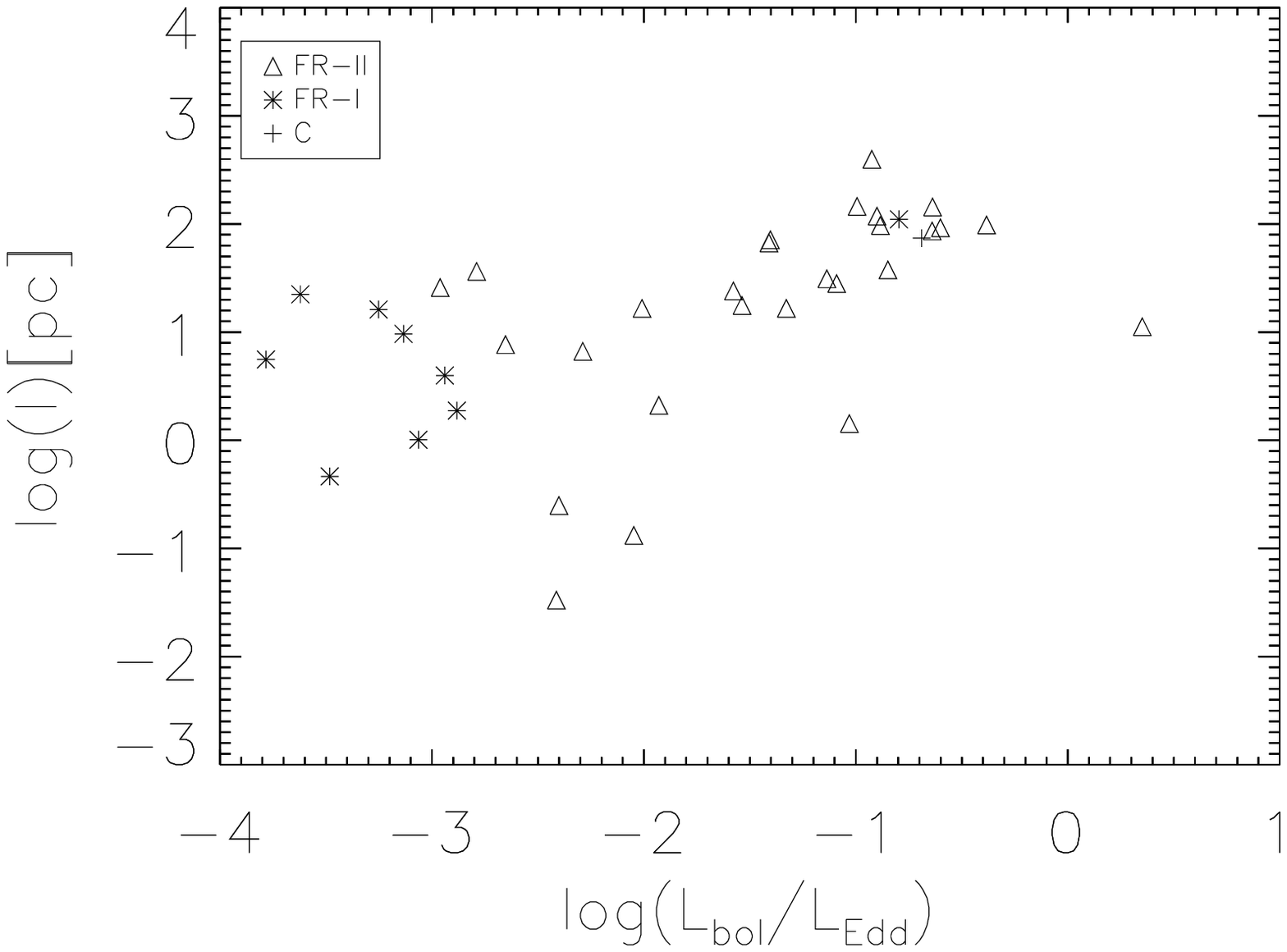}
	\end{minipage}%
	\vspace*{0.5cm}
	\centering
	\caption{Left: the distribution of the projected linear size of the jets for 37 sources measured in this work (see text for details). Right: the linear size vs. the Eddington ratio.  \label{fig:size}}	
\end{figure}
\begin{figure}[h]
	\begin{minipage}[t]{0.5\linewidth}
		\centering
		\includegraphics[width=70mm]{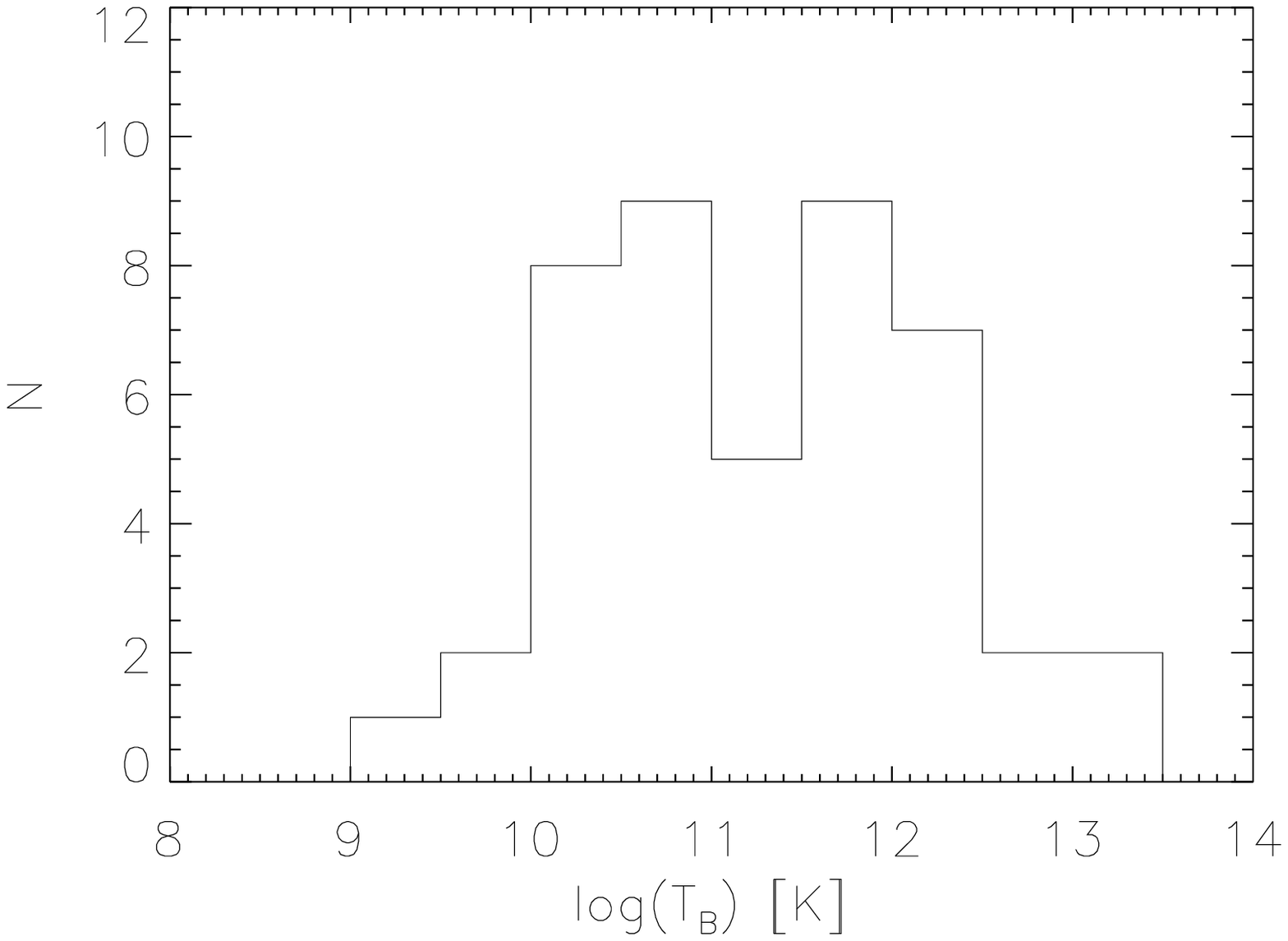}		
	\end{minipage}%
	\begin{minipage}[t]{0.5\textwidth}
		\centering
		\includegraphics[width=70mm]{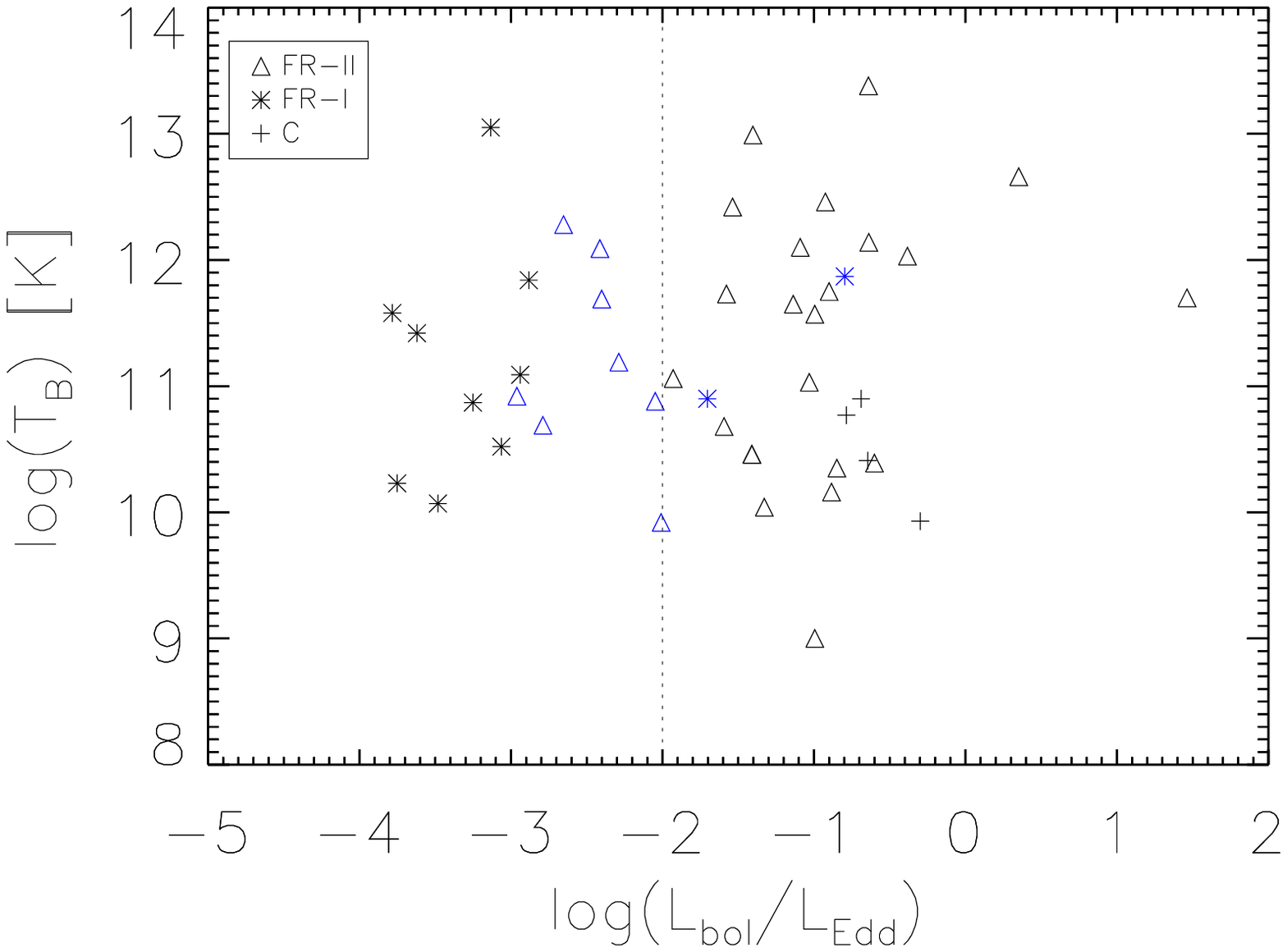}
	\end{minipage}%
	\vspace*{0.5cm}
	\centering
	\caption{Left: the distribution of the brightness temperature of VLBA radio core. Right: the core brightness temperature versus the Eddington ratio. The dotted line is $L_{\rm bol}/L_{\rm Edd}=0.01$. \label{fig:T_B}}	
\end{figure}
\begin{figure}[h]
	\begin{minipage}[t]{0.5\linewidth}
		\centering
		\includegraphics[width=65mm]{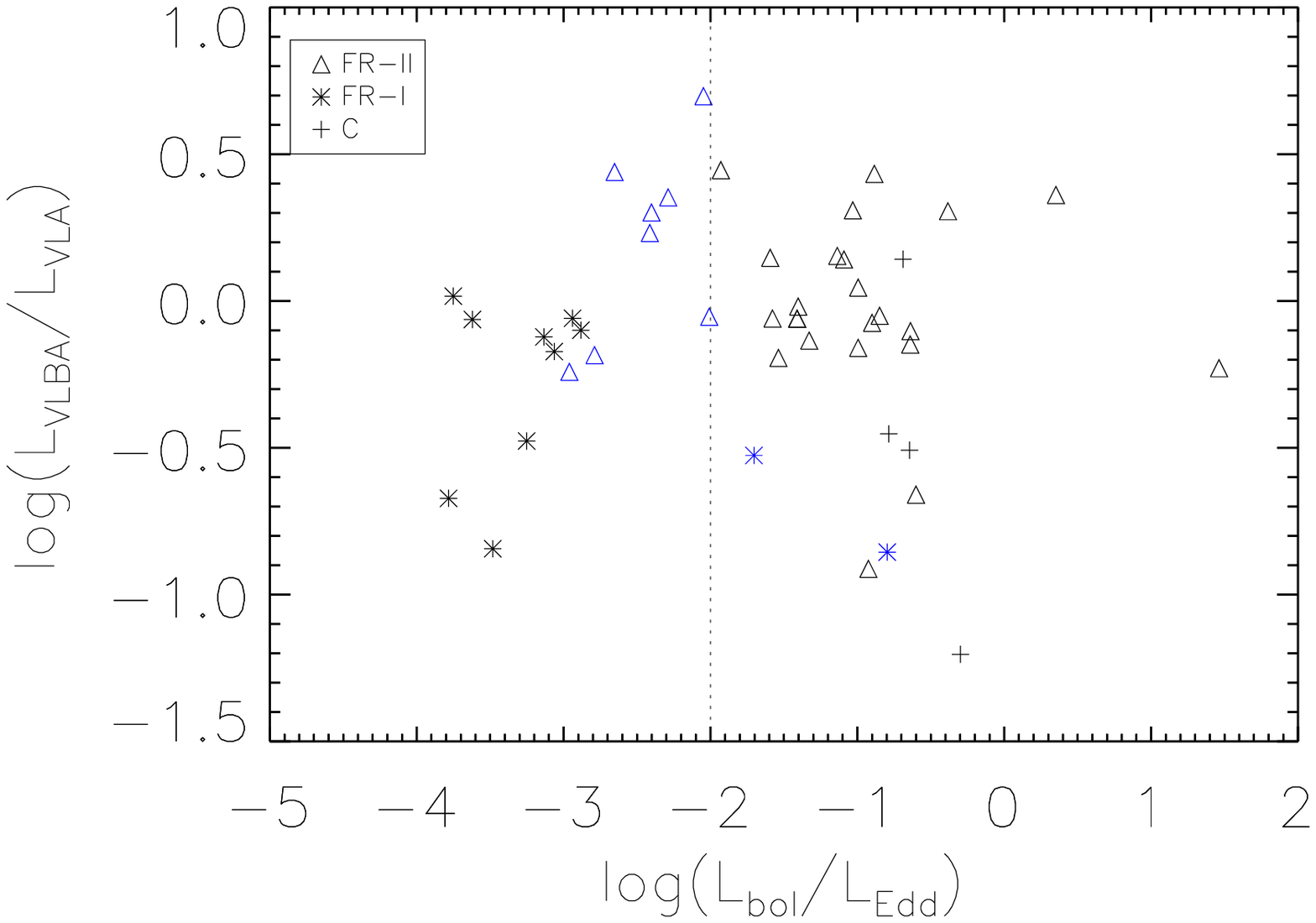}		
	\end{minipage}%
	\begin{minipage}[t]{0.5\textwidth}
		\centering
		\includegraphics[width=65mm]{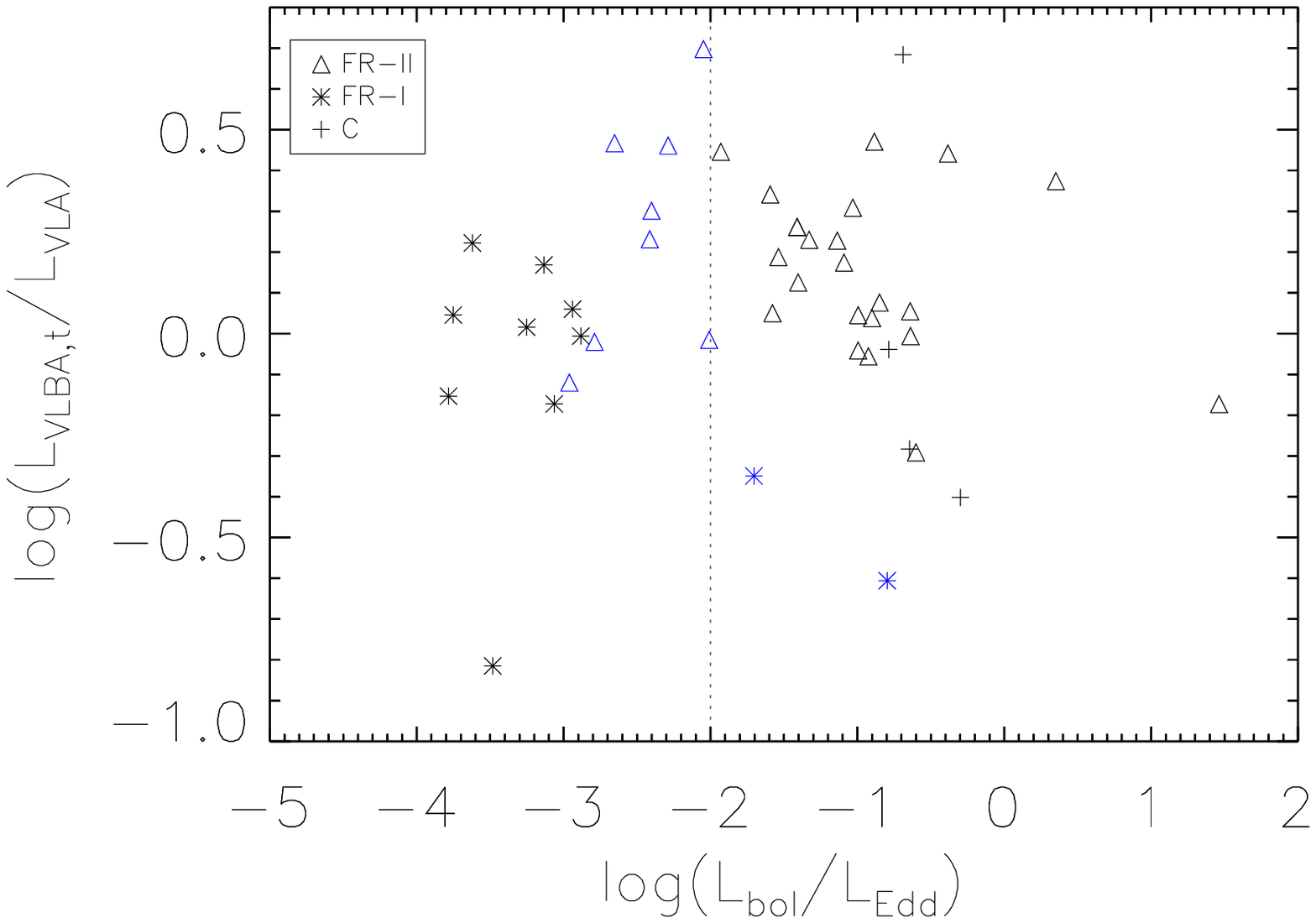}
	\end{minipage}%
	\vspace*{0.2cm}
	\centering
	\caption{Left: the flux ratio of VLBA core to VLA core at 5 GHz versus the Eddington ratio. Right: the flux ratio of VLBA total to VLA core at 5 GHz versus the Eddington ratio. The dotted line is $L_{\rm bol}/L_{\rm Edd}=0.01$.  \label{fig:compac}}	
\end{figure}
\begin{figure}[h]
	\begin{minipage}[t]{0.5\linewidth}
		\centering
		\includegraphics[width=70mm]{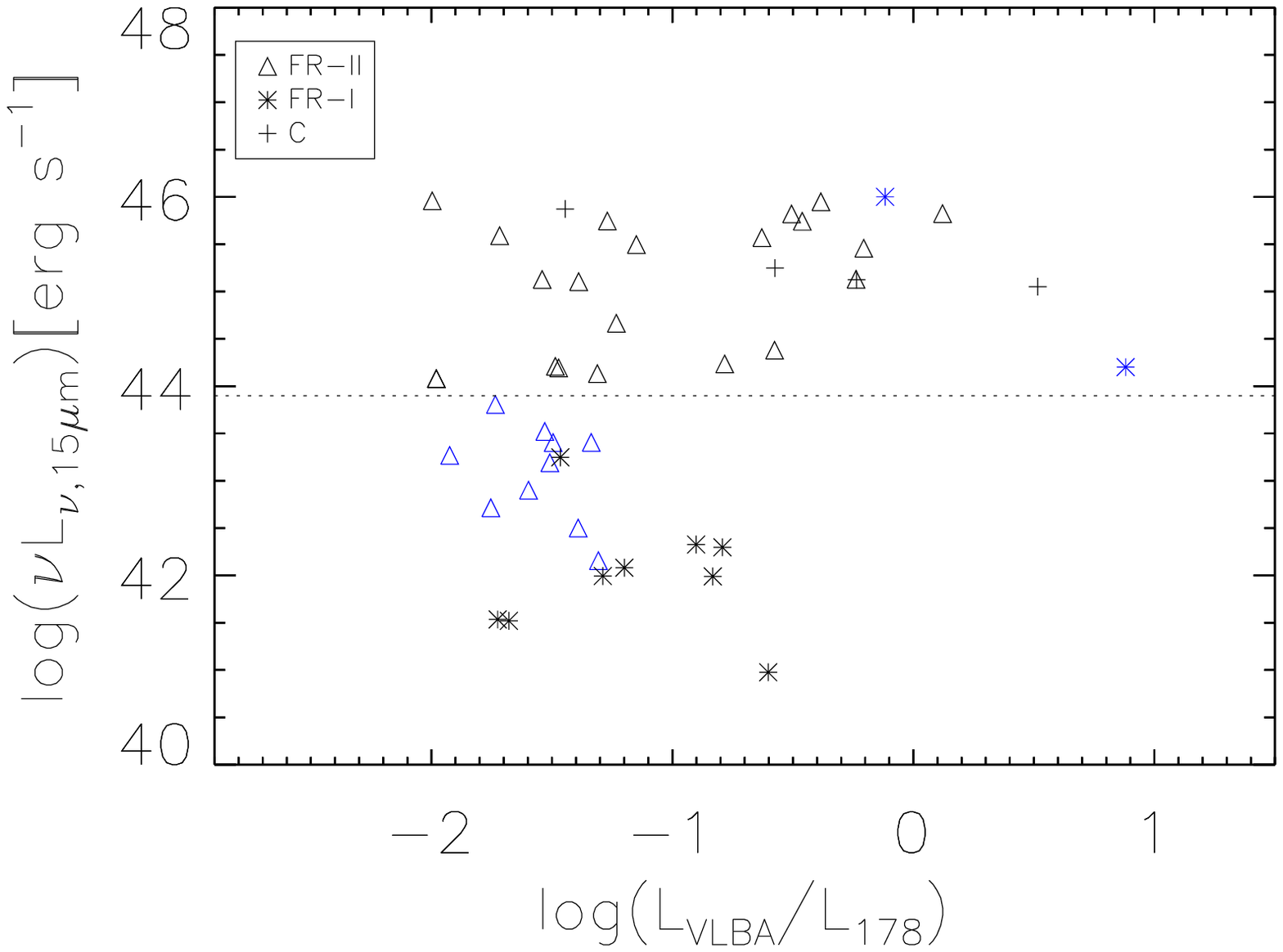}		
	\end{minipage}%
	\begin{minipage}[t]{0.5\textwidth}
		\centering
		\includegraphics[width=70mm]{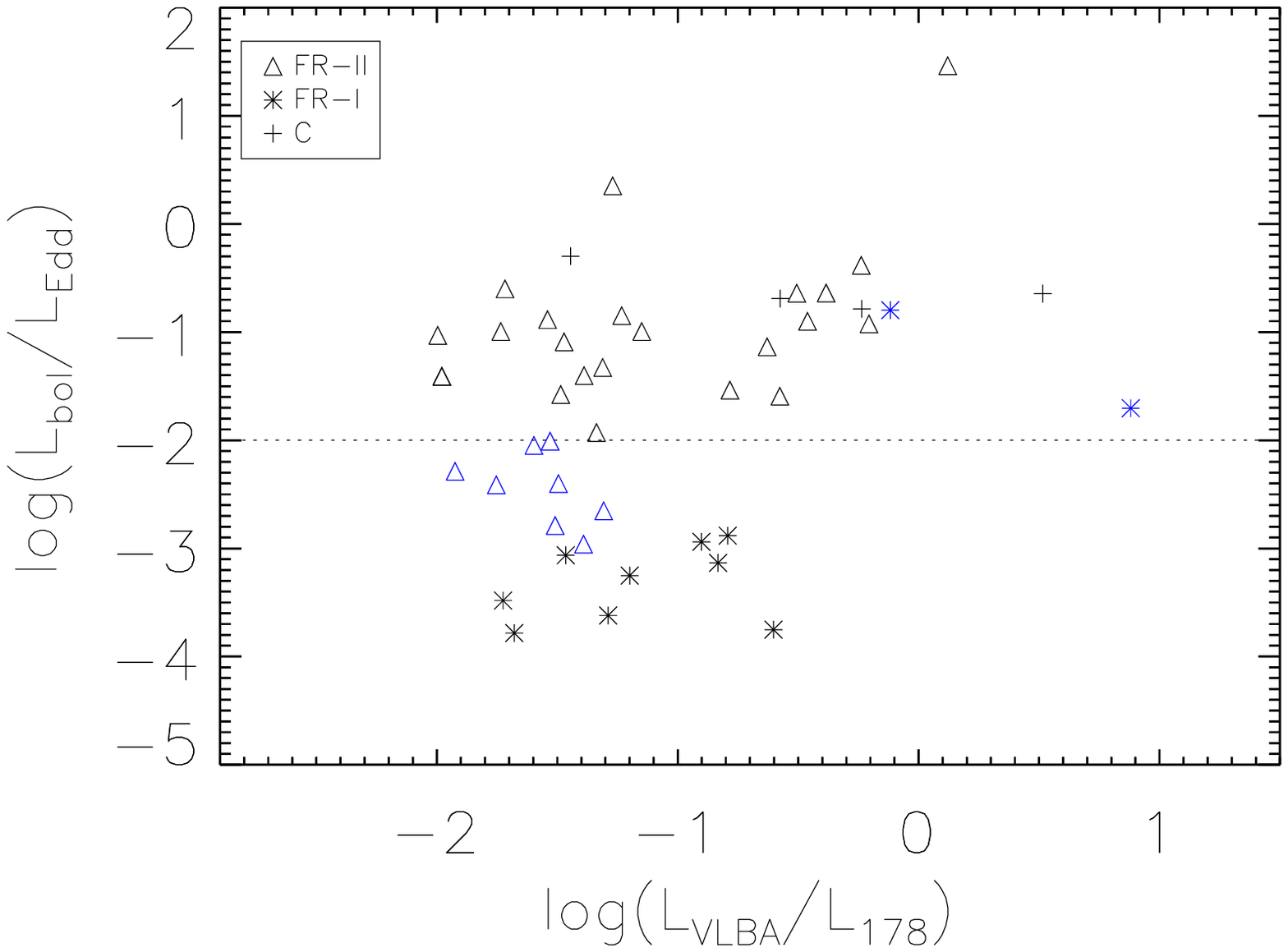}
	\end{minipage}%
	\vspace*{0.5cm}
	\centering
	\caption{Left: The relation of the luminosity at 15 $\rm \mu m$ and the ratio of VLBA core at 5 GHz to the 178 MHz luminosity. The dotted line is $\nu L_{\nu,15 \mu m} = 8 \times10^{43} \rm erg ~s^{-1}$. Right: The relation of the Eddington ratio and the ratio of VLBA core at 5 GHz to the 178 MHz luminosity. The dotted line is $L_{\rm bol}/L_{\rm Edd}=0.01$. \label{fig:corelobe}}	
\end{figure}

\label{lastpage}
\end{document}